\documentclass[preprint,12pt]{elsarticle}

\usepackage{amssymb}
\usepackage{amsmath}
\usepackage{subfigure}
\journal{Computer Physics Communications}

\begin{document}

\begin{frontmatter}



\title{A three-dimensional domain decomposition method for large-scale DFT electronic structure calculations}


\author[jaist,issp]{Truong Vinh Truong Duy \corref{cor1}\fnref{label2}}
\ead{duytvt@jaist.ac.jp}
\author[jaist]{Taisuke Ozaki \corref{cor2}\fnref{label2}}
\ead{t-ozaki@jaist.ac.jp}
\address[jaist]{Research Center for Simulation Science, Japan Advanced Institute of Science and Technology (JAIST), 1-1 Asahidai, Nomi, Ishikawa 923-1292, Japan}
\address[issp]{Institute for Solid State Physics, The University of Tokyo, Kashiwanoha 5-1-5, Kashiwa, Chiba 277-8581, Japan}

\cortext[cor1]{Corresponding author. Tel.: +81 761 51 1987}
\cortext[cor2]{Principal corresponding author. Tel.: +81 761 51 1582. }
\fntext[label2]{These authors contributed equally to this work.}

\begin{abstract}
With tens of petaflops supercomputers already in operation and exaflops machines expected to appear within the next 10 years, efficient parallel computational methods are required to take advantage of such extreme-scale machines.
In this paper, we present a three-dimensional domain decomposition scheme for enabling large-scale electronic calculations based on density functional theory (DFT) on massively parallel computers.
It is composed of two methods: (i) atom decomposition method and (ii) grid decomposition method. In the former, we develop a modified recursive bisection method based on inertia tensor moment to reorder the atoms along a principal axis so that atoms that are close in real space are also close on the axis to ensure data locality. The atoms are then divided into sub-domains depending on their projections onto the principal axis in a balanced way among the processes. In the latter, we define four data structures for the partitioning of grids that are carefully constructed to make data locality consistent with that of the clustered atoms for minimizing data communications between the processes. We also propose a decomposition method for solving the Poisson equation using three-dimensional FFT in Hartree potential calculation, which is shown to be better than a previously proposed parallelization method based on a two-dimensional decomposition in terms of communication efficiency.  
%We also give a detailed analysis on the amount of communication in the Hartree potential calculation using FFT, and apply the best pattern in our grid decomposition method, which is shown to be up to 77.8\% better than a previously proposed parallelization method based on a two-dimensional decomposition in terms of communication efficiency. 
For evaluation, we perform benchmark calculations with our open-source DFT code, OpenMX, paying particular attention to the $O(N)$ Krylov subspace method. The results show that our scheme exhibits good strong and weak scaling properties, with the parallel efficiency at 131,072 cores being 67.7\% compared to the baseline of 16,384 cores with 131,072 diamond atoms on the K computer.  
\end{abstract}


\end{frontmatter}

% \linenumbers

\section{Introduction}
\label{Introduction}
Density functional theory (DFT) \cite{Hohenberg:1964zz,Kohn:1965zzb,Hedin:1965zza} is widely regarded as one of the most powerful and popular methods available in computational materials science simulations. It has been applied to various systems of interest, from physics, chemistry, and materials science to biology, for exploring a wide range of material and biological properties, such as structural and optical properties, electric transport, chemical reactions, etc. Although the Kohn-Sham (KS) method offers a feasible approach to performing the calculation of the ground state properties of $N$-body systems, its scaling property of $O(N^3)$ has posed great challenges in large-scale calculations.
There has been considerable effort to realize large-scale DFT calculations that can be roughly classified into two major approaches: first, improve and implement highly efficient parallelization of conventional methods, and second, develop better-than-cubic scaling methods, ideally linear scaling methods.   

In the first approach, real-space methods are well-established and significant progress in developing practical real-space methods and implementations has been observed. Recently, a real-space DFT code has achieved an unprecedented performance on the K computer \cite{hasegawa2011first}. Real-space implementations of DFT applying finite difference methods \cite{saad2010numerical} and finite element methods \cite{PhysRevB.52.5573,PhysRevB.54.7602,JPSJ.67.3844,pask2005finite} have also been extensively developed. On the other hand, in the second approach, a number of linear scaling methods have been proposed. Several survey reports on linear scaling methods can be found in literature \cite{goedecker1999linear,bowler2002recent,goedecker2003linear,bowler2012methods,skylaris2005introducing}. A popular and straightforward approach to linear scaling is the divide and conquer (DC) method \cite{yang1991direct,yang1995density,khandogin2003insights}. Recursion method proposed by Ozaki et al. in \cite{ozaki1999bond,ozaki2000block,ozaki2001convergent} is another approach. Other linear scaling methods include the early work presented in \cite{kohn1996density}, methods based on Wannier functions  \cite{stephan1998order,ordejon1995linear,mauri1993orbital,xiang2006linear} and density matrix \cite{rudberg2011assessment,jordan2005comparison,PhysRevB.47.10895,PhysRevB.47.10891}, and methods for tight-binding systems \cite{goedecker1994efficient,goedecker1995tight} and for metallic and non-metallic systems \cite{krajewski2005stochastic,krajewski2006linear}.

Our main concern in this paper is the development of an efficient three-dimensional domain decomposition method that is applicable to both conventional methods and linear scaling methods to enable large-scale electronic calculations based on DFT on massively parallel computers. In fact, our linear scaling DFT method based on Krylov subspace method has been developed and introduced in \citep{ozaki2006n}. Even though there exist a number of parallel implementations of linear scaling DFT \cite{bowler2010calculations,artacho2008siesta,haynes2006onetep} and domain decomposition methods based on the space-filling curve technique \cite{brazdova2008automatic,challacombe2000general}, these methods may suffer from difficulty in applying to anisotropic structures.
%have limited applicability to anisotropic structures. %their demonstrations for parallel performance are limited to just several hundreds to a few thousand cores. 
Considering the fact that we are now in the petaflops era with several supercomputers having a peak performance of tens of petaflops and exaflops machines are expected to appear within the next 10 years with millions of cores, generally applicable and efficient methods are required to take advantage of such extreme-scale machines and solve a wide variety of problems. In parallelizaton of linear scaling methods, as our implementation is also based on atomic orbitals \cite{hehre1986ab} and their linear combination \cite{sanchez1997density,ozaki2003variationally}, a known problem arises due to the use of localized atomic basis functions with different spatial cut-off radius for different elements. The irregular sparse structure of the resultant matrices make their parallel multiplication implementation become more complicated, since they require careful mapping of the sparse matrix elements to processing cores, unlike finite element methods where the matrices are regular. In addition, the domain decomposition method should hold the locality of clustered atoms in real space and assign nearby atoms to processors to minimize inter-node communications. Also, it should be applicable to any distribution patterns of atoms in real space.

To address the issues, in this paper we propose a domain decomposition scheme, which is actually composed of two methods: one for decomposing the atoms and the other for partitioning the grids among the processes. In the former, we develop a modified recursive bisection method based on inertia tensor moment to reorder the atoms along a principal axis so that atoms that are close in real space are also close on the axis. Depending on their positions on the axis, the atoms are then divided into sub-domains in a balanced way with data locality conserved. The method should be able to work well with any number of atoms and processes. 
%, and assign each sub-region to one process. 
Meanwhile, in the latter, we introduce four data structures for the grid partitioning that are carefully constructed to make data locality consistent with that of the clustered atoms, resulting in the minimization of data communications between the processes. 
%facilitate the calculations of charge density and other potentials.
We furthermore propose a decomposition method for solving the Poisson equation using three-dimensional (3D) FFT in Hartree potential calculation, which is proven to be more communication efficient than one- and two-dimensional decomposition methods. 
%A detailed analysis on the amount of communication using FFT is given to employ the best calculation order pattern compared with two other decomposition methods. 
We also let the processes perform on-the-fly communications to reduce the memory usage. 

The remainder of the paper is organized as follows. Section 2 describes briefly the research background, including the KS equation and our linear scaling Krylov subspace method. The domain decomposition methods for atoms and grids are presented in Section 3. Section 4 details implementation issues and the calculation flowchart. Section 5 analyzes benchmark results with a focus on linear scaling, and strong and weak scaling properties. Finally, we conclude our study in Section 6. 

%\section{{\it Ab initio} electronic structure calculations}
\section{Background}
%\label{Electronic-structure }
For the sake of completeness and self-explanation, we briefly introduce the background of DFT by way of the well-known KS equation, and give an overview of our linear scaling Krylov subspace method, while highlighting data structures related to our parallelization scheme. %s for solving the equation.  

\subsection{Kohn-Sham equation}
In principle, DFT is a modeling method to determine the electronic ground state of $N$-body systems of atoms and molecules based on functionals of electron density. Although DFT may have its root date back to the Thomas-Fermi model and Slater's X$\alpha$ method derived from a simplification of the Hartree-Fock method \cite{thomas1927calculation,slater1951simplification}, its theoretical foundation was only established firmly by the Hohenberg-Kohn theorems introduced in the 1960s \cite{Hohenberg:1964zz}. The resulting KS equation is typically represented in the form of an eigenvalue equation as below \cite{parr1994density,martin2004electronic}.

\begin{equation}
%\left [ -\frac{1}{2} \triangledown^2  + \upsilon _{eff}(r)\right]\phi _{i}(r)=\varepsilon _{i}\phi _{i}(r)
%\left [ -\frac{1}{2} \triangledown^2  + \upsilon _{eff}(r)\right] \psi _{k}(r)=\varepsilon _{k} \psi _{k}(r)
\hat{H}_{\mathrm{KS}}\phi _{\nu }(\textbf{\textit{r}})=\varepsilon _{\nu }\phi _{\nu }(\textbf{\textit{r}}),
\end{equation}
where $\hat{H}_{\mathrm{KS}}$ is the KS Hamiltonian, and $\varepsilon _{\nu }$ is the orbital energy of the corresponding KS orbital $\phi _{\nu }(\textbf{\textit{r}})$. Consequently, the density $n(\textbf{\textit{r}})$ of the system is defined as: 
\begin{equation}
n(\textbf{\textit{r}})=2\sum_{i=1}^{\mathrm{occ.}}\phi _{\nu }^{*}(\textbf{\textit{r}})\phi _{\nu }(\textbf{\textit{r}}),
\end{equation} 
where the factor of 2 is due to spin multiplicity, and $\nu $ runs for all the occupied states. Although our discussion is limited within non-spin polarization, the generalization to spin polarization is straightforward.

In our implementation, we utilize atomic orbital basis functions, written as the multiplication of a spherical harmonic $Y_{l}^{m}$ and a radial function $R$:
\begin{equation}
\label{basis-function}
\phi_{\nu }(\textbf{\textit{r}}) = Y_{l}^{m}(\widehat{\textbf{\textit{r}}})R(r).
\end{equation} 
The basis functions have been carefully constructed and optimized for convergence and reduction of computational effort with high accuracy, as well as memory usage \cite{ozaki2003variationally}. Most notably, the use of atomic-like orbitals perfectly matches the idea of linear scaling methods. They are strictly confined within a sphere and stored in a special structure described later in Section \ref{data-structure}. In the same section, another structure for storing all the grid points inside a parallelepipedon defined by the basis functions of all atoms distributed to the same process in the parallel implementation will also be presented. 

The Hamiltonian is expressed as: 
\begin{equation}
\hat{H}_{\mathrm{KS}}= -\frac{1}{2} \triangledown^2  + \upsilon _{\mathrm{eff}}(\textbf{\textit{r}}),
\end{equation}
where $\upsilon _{\mathrm{eff}}(\textbf{\textit{r}})$ is the Kohn-Sham effective potential, given by:
\begin{equation}
\label{eq-KS-potential}
\upsilon _{\mathrm{eff}}(\textbf{\textit{r}})=\upsilon _{\mathrm{ext}}(\textbf{\textit{r}})+\upsilon _{\mathrm{Hartree}}(\textbf{\textit{r}})+\frac{\delta E_{\mathrm{xc}}[n]}{\delta n(\textbf{\textit{r}})}.
\end{equation}
Here, the first term, $\upsilon _{\mathrm{ext}}(\textbf{\textit{r}})$, is the external potential due to the interaction between electrons and nuclei and other external fields. 
The second term $\upsilon _{\mathrm{Hartree}}(\textbf{\textit{r}})$ is the classical Hartree (Coulomb) interaction of the electron density interacting with itself, given by 
\begin{equation}
\label{Hartree-potential}
\upsilon _{\mathrm{Hartree}}(\textbf{\textit{r}})=\int \frac{n(\textbf{\textit{r}})}{\left | \textbf{\textit{r}}-\textbf{\textit{r}}^{'} \right |}d\textbf{\textit{r}}^{'}.
\end{equation}
In our implementation, the Hartree potential is found by solving the Poisson equation using FFT. In parallel implementations, the grid points must be decomposed to the processes in a proper fashion to minimize the communication amounts incurred during the calculation. We will introduce our grid decomposition method and compare with other methods in terms of communication efficiency in Section \ref{grid-decomposition}.  

The last term in Eq. (\ref{eq-KS-potential}) is the exchange-correlation potential, where $E_{\mathrm{xc}}$ is the corresponding exchange-correlation energy that is the unknown in this KS approach. A number of methods have been proposed to yield approximations to this exchange-correlation energy, for instance, local density approximation (LDA) and generalized gradient approximation (GGA) \cite{parr1994density,martin2004electronic}. We will construct a particular structure for the calculation of this potential in Section \ref{data-structure}.

The KS equation turns out to be in the form of one-body equations under the effective potential. Combined with the methods for approximating $E_{\mathrm{xc}}$, the equation can be solved self-consistently with a resulting ground state density. Then the corresponding energy can be determined:    
\begin{equation}
\label{total-energy}
E_{\mathrm{KS}}[n]=T_{\mathrm{s}}+\int n(\textbf{\textit{r}})\upsilon _{\mathrm{ext}}(\textbf{\textit{r}})d\textbf{\textit{r}}+E_{\mathrm{Hartree}}[n]+E_{\mathrm{xc}}[n],
\end{equation}
where $T_{\mathrm{s}}$ is the KS kinetic energy of non-interacting electrons, also expressed in terms of the KS orbitals:
\begin{equation}
T_{\mathrm{s}}=-\sum_{\nu =1}^{\mathrm{occ.}}\int \phi _{\nu }^{*}\bigtriangledown ^{2}\phi _{\nu }d\textbf{\textit{r}}.
\end{equation}

\subsection{Linear scaling Krylov subspace method}
%\subsection{Motivation}

%\section{Linear scaling Krylov subspace method}

\begin{figure}[htb]
\centering
\includegraphics[scale=0.5,trim=0cm 0cm 0cm 0cm]{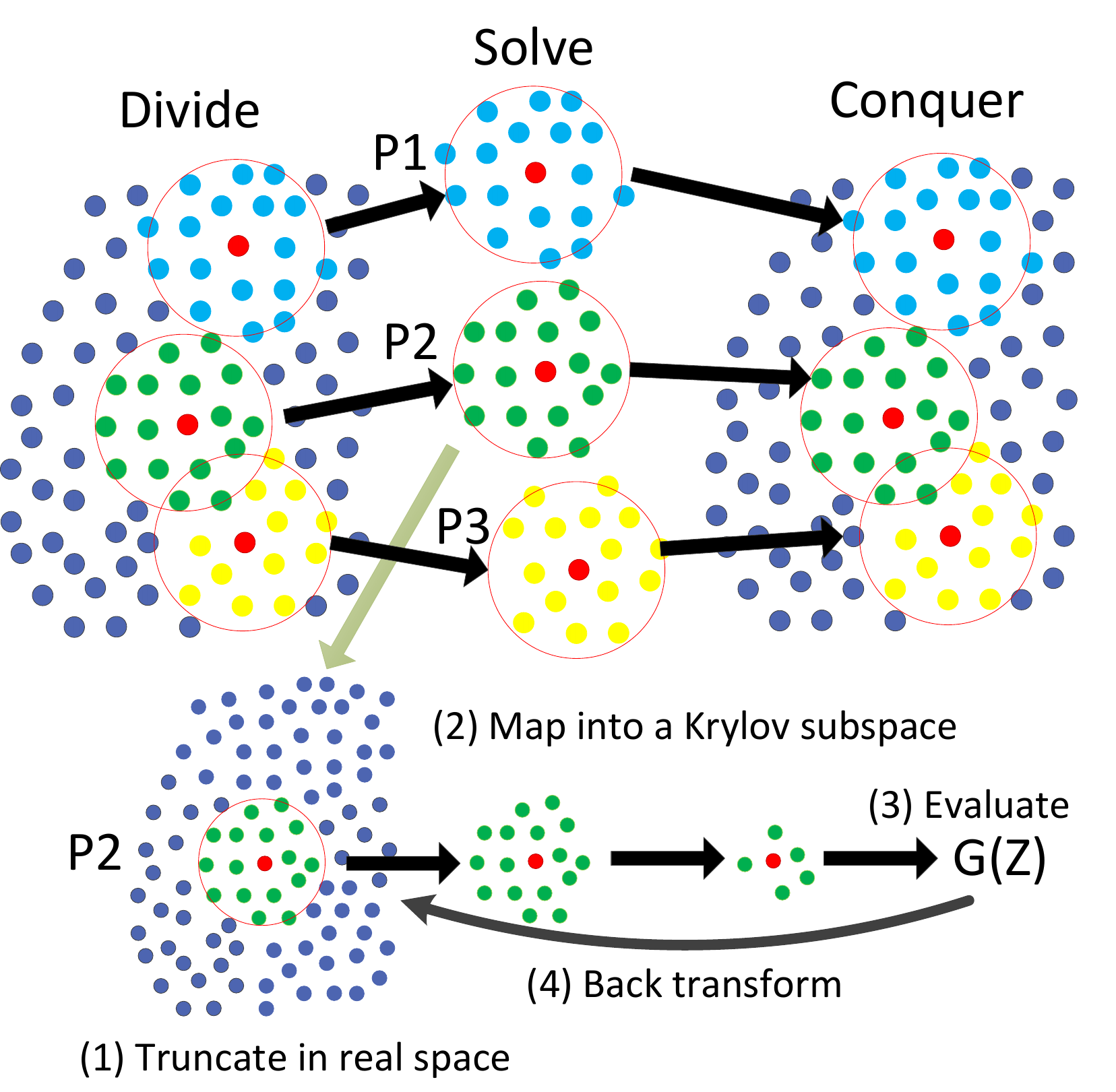}
\caption{Linear scaling Krylov subspace method.}
\label{fig-krylov-method}
\end{figure}

%This section gives a brief introduction to the linear scaling Krylov subspace method \citep{ozaki2006n}, which is a unified method of the divide and conquer (DC) method and recursion method. The DC method has been proven good for covalent systems but not for metals because of the large size of a truncated cluster due to a direct diagonalization technique, while the recursion method is known for its numerical instability in SCF calculations. They are founded on the fact that the charge density $n$ can be given by summing up the local charge density $n_i$ of each site, which is actually evaluated with the density matrix $\rho$:
The basic idea behind the Krylov subspace method \citep{ozaki2006n} is to combine the DC method and the recursion method based on the Green's function. The DC method has been proven good for covalent systems but not for metals due to the difficulty of direct diagonalization for large truncated clusters, while the recursion method is known for its numerical instability in SCF calculations. The three methods are founded on the same fact that the charge density $n$ can be given by summing up the local charge density $n_i$ of each site, which is actually evaluated with the density matrix $\rho$:
\begin{equation}
\label{charge-density}
n(\textbf{\textit{r}})=2\sum_{i\alpha ,j\beta }\phi_{i\alpha }(\textbf{\textit{r}})\phi_{j\beta }(\textbf{\textit{r}})\rho _{i\alpha ,j\beta}
=2\sum_{i}\left (\sum_{\alpha ,j\beta }\phi_{i\alpha }(\textbf{\textit{r}})\phi_{j\beta }(\textbf{\textit{r}})\rho _{i\alpha ,j\beta}  \right )
=2\sum_{i}n_{i}(\textbf{\textit{r}}),
\end{equation}
where $i$ is the site index, and $\alpha$ is the orbital index of the basis functions.

The density matrix $\rho$ is in turn calculated by the one-particle Green's function:
\begin{equation}
\label{density-matrix}
\rho _{i\alpha ,j\beta}=-\frac{1}{\pi }\mathrm{Im}\int G_{i\alpha ,j\beta}(E+i0^{+})f\left ( \frac{E-\mu }{k_{B}T} \right )dE,
\end{equation}
where $f(x)\equiv 1/[1+\mathrm{exp}(x)]$ is the Fermi function, and $0^+$ is a positive infinitesimal.

The problem now turns to how to evaluate the Green's function in linear scaling time, and hence each method has its own approach. The Krylov subspace method can be seen as a DC method defined in a Krylov subspace for reducing the size of the truncated clusters, as the Krylov subspace is much smaller than the original vector space. In this method, the total Green's function can be approximated by the sum of the local Green's functions associated with the central atom in each truncated cluster. 

Figure \ref{fig-krylov-method} demonstrates how the Krylov subspace method works in practice.
First, in the Divide step, the original system is decomposed into smaller truncated clusters using a physical truncation scheme, described later in Section \ref{Implementation}. The truncated clusters are assigned to different processes for processing in parallel.  
Then, in the Solve step, the Krylov subspace for each truncated cluster is generated independently on each process without any communications. After that the Green's function associated with each truncated clusters is evaluated, and the back transform is performed to get the Green's functions represented by the original basis functions for calculating the density matrix $\rho$ which is then found by Eq. (\ref{density-matrix}), followed by the local charge density $n_i$ by Eq. (\ref{charge-density}). 
Finally, in the Conquer step, the total charge density $n$ is given by combining all the local charge densities $n_i$ (Eq. (\ref{charge-density})). The density is treated by regular mesh, and we will therefore introduce the grid decomposition method with four data structures for storing data on the grid points in Section \ref{grid-decomposition}.
The Krylov subspace method has been proven to be efficient, robust, could converge rapidly for both metals and insulators, and has been recently applied to a large-scale molecular dynamics simulation for electrochemical systems \citep{ohwaki2012large} and structural prediction of precipitates of a carbide in bcc iron \citep{Sawada2012}. Interested readers are referred to \citep{ozaki2006n} for details. 
%The original system is therefore first truncated into several clusters, which can be solved independently, and the results are then combined together.  

%Combination of divide and conquer method () and recursion method based on Lanczos algorithm and the Green’s function (numerical instability during SCF calculations) -> DC method defined in a Krylov subspace. The Krylov subspace Uk < original vector space Uc < Uf. 

\section{Domain decomposition methods}
\label{Methods}
Domain decomposition should be performed in three dimensions, especially for large-scale calculations with a large number of processes, considering the scaling properties of communication amount and memory usage. One- and two-dimensional decomposition methods are  only appropriate for up to a certain number of processes, and when that number exceeds the limit of one or two dimensions, the amount of communication and memory usage will stay unchanged. In contrast, three-dimensional methods can realize much larger scales of calculations, as they are perfectly suited to the three-dimensional nature of systems' structure.   
In this section, we present our three-dimensional domain decomposition scheme, consisting of two methods: one for the atom level and the other for the grid level in three-dimensional space. 

\subsection{Three-dimensional atom decomposition method}
\subsubsection{Requirements}
In parallel implementations, atom decomposition among the processes is a very important first task that has a considerable impact on the overall performance.  
We consider a 3D domain decomposition method for atoms in a system, which should hold the locality of clustered atoms in real space with the following requirements.  
\begin{itemize}
\item
{\bf Approximately the same computational amount}: Each sub-domain should have approximately the same computational amount. This requirement is obvious for ensuring a good load balance among the processes. As atoms of different elements have different processing time, each atom should have a corresponding weight reflecting the amount of computation. The only exception is the case of the first MD step or geometry optimization step, where all the atoms have the same weight of 1. 
\item
{\bf Locality}: Nearby atoms should be grouped together and assigned to the same process. This will reduce the amount of communication and assists the development of linear scaling methods, which are also essentially based on the idea of locality.   
\item
{\bf Applicable to any number of atoms and processes}: Realistic systems can have any number of atoms, and can be solved by any number of processing cores available to maximize resource usage. Hence, in order to treat them effectively, the method should be able to work with any number of atoms and processes.   
\item
{\bf Applicable to any distribution pattern of atoms}: Some specific systems have most atoms distributed around the center of mass, while some have atoms spread uniformly in the space, and others may have a majority of atoms lie along just one main dimension. Such irregular distributions of atoms should be considered in the method.
\item
{\bf A well-defined algorithm}: To be clear and easy to follow. 
\end{itemize}

\begin{figure}[htb]
\centering
\includegraphics[scale=0.6,trim=0cm 0cm 0cm 0cm]{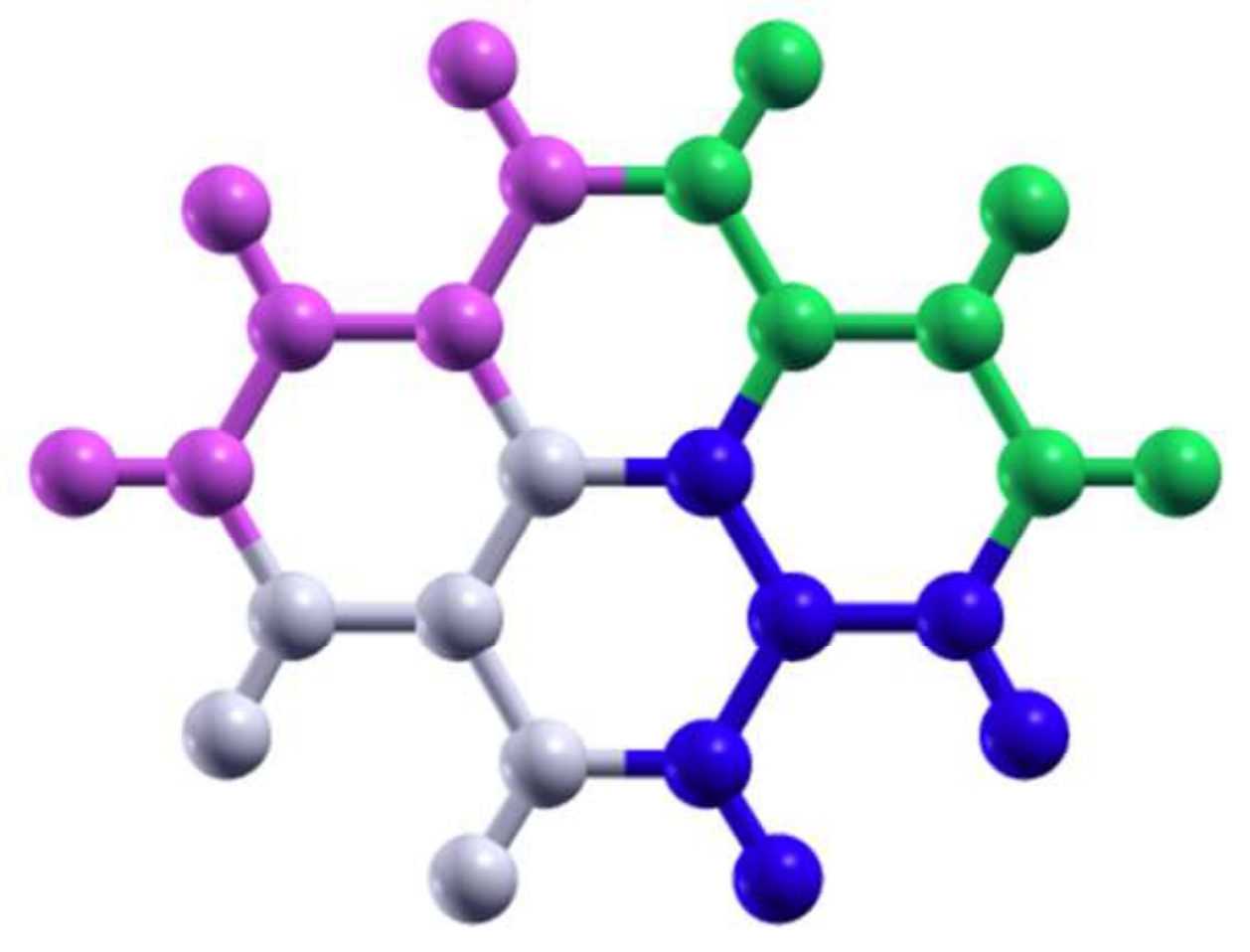}
\caption{An example of domain decomposition for atoms with each color representing a sub-domain.}
\label{fig-domain-atom}
\end{figure}

Figure \ref{fig-domain-atom} exhibits an expected domain decomposition result for a system of 26 atoms, in 2D for the sake of simplicity, where nearby atoms are nearly equally allocated into four sub-domains. 
Given the requirements, a proper 3D domain decomposition can be performed based on two ideas: modified recursive bisection and inertia moment tensor, which are presented subsequently.  

\subsubsection{Modified recursive bisection method}
The modified recursive bisection method is a useful tool to decompose the system with any number of processes and atoms, unlike the conventional recursive bisection method that is restricted to numbers equal to a power of 2 \citep{salmon1991parallel}. The method involves constructing a binary tree, and each tree node has a number representing the number of processes treating the domain under that node. Leaf nodes have a number of processes valued at 1, and the number of leaf nodes is equal to the number of processes. In practice, the method works as follows.

First, the original domain is divided into two sub-domains, each having approximately the same number of processes, which is about half of that of the original domain. Each sub-domain is then further divided into two, again with each having nearly the same number of processes. The division is repeated until the number of processes becomes 1. Once the division procedure has been completed, there are as many leaf nodes as processes.  

\begin{figure}[htb]
\centering
\includegraphics[scale=0.8,trim=0cm 0cm 0cm 0cm]{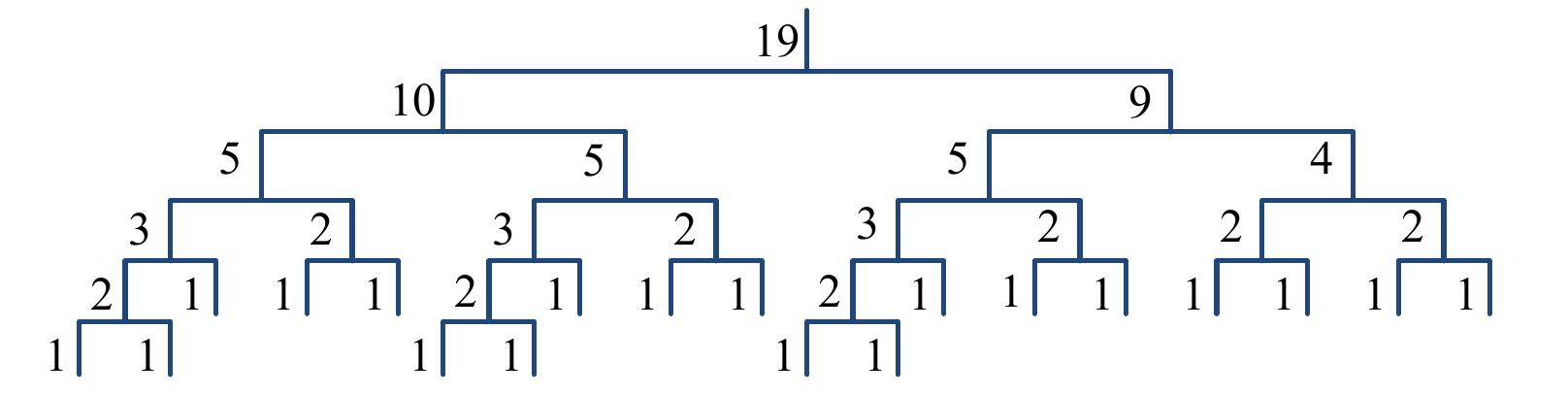}
\caption{Modified recursive bisection method.}
\label{fig-modified-recursive}
\end{figure}

Figure \ref{fig-modified-recursive} shows an example of the method where the number of processes is 19. It is also important to note that in the conventional recursive bisection method limited to the number of processes equal to a power of 2, the bisection is made so that equal numbers can be assigned to each sub-domain. However, the modified method bisects with weights as shown in the figure.

\subsubsection{Inertia tensor moment}
\begin{figure}[htb]
\centering
\includegraphics[scale=0.8,trim=0cm 0cm 0cm 0cm]{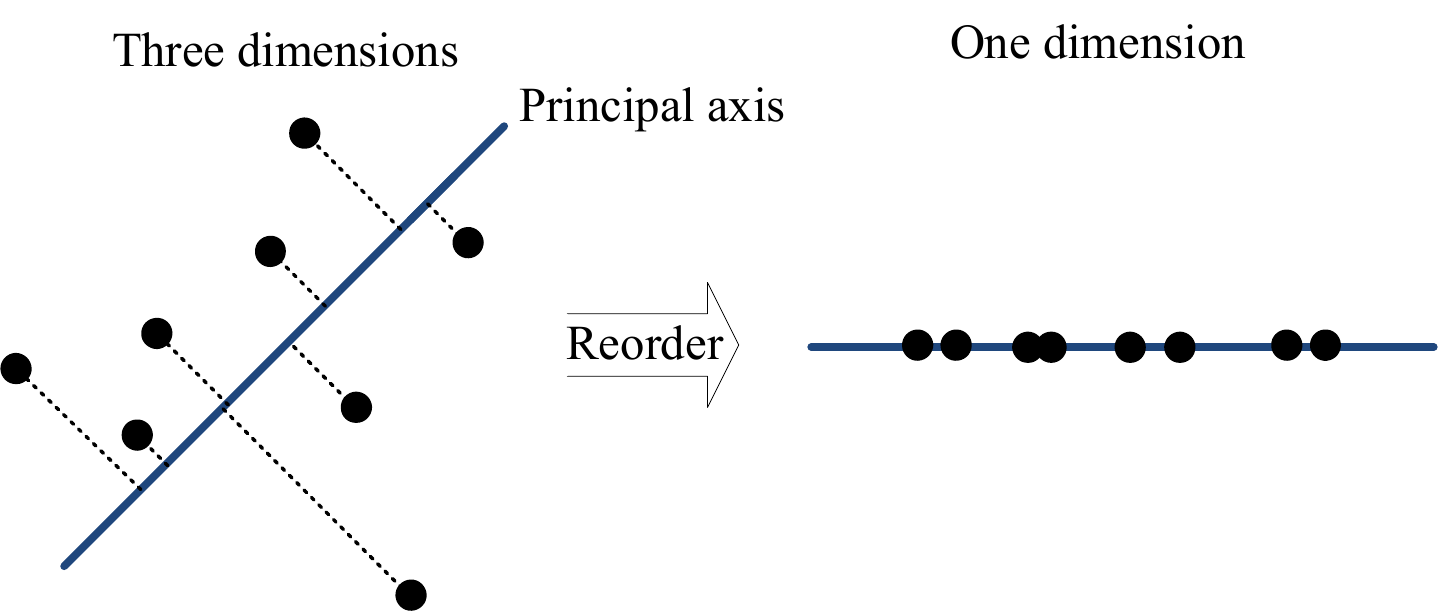}
\caption{Reordering of atoms by a principal axis found by the inertia tensor moment.}
\label{fig-inertia-tensor}
\end{figure}

If the modified recursive bisection method meets the requirement to work with any number of processes and atoms, the use of an inertia moment tensor for the bisection in the method ensures the locality of the domain decomposition. The inertia moment tensor is applied to calculate a principal axis for each sub-domain during the decomposition process described above. Then atoms in the corresponding sub-domain are reordered from three dimensions to one dimension by projecting them onto the principal axis. This reordering will also make it much easier to divide the atoms into two sub-domains to fit the structure of the binary tree. The inertia tensor moment approach is expected to work well, as atoms that are close in real space are also close on the principal axis yielded by the inertia tensor. Figure \ref{fig-inertia-tensor} illustrates the idea of 3D-to-1D reordering of atoms. 

Mathematically, the equation of the principal axis can be written as:
\begin{equation}
\begin{matrix}
x=C_{x}+a_{x}t \\ 
y=C_{y}+a_{y}t\\ 
z=C_{z}+a_{z}t,
\end{matrix}
\end{equation}
where $(C_{x}, C_{y}, C_{z})$ is the center of mass of the system, defined as the average of the atoms' positions, $x_{i}$, weighted by their weights, $w_{i}$, with the total number of atoms, $N_a$:
\begin{equation}
\label{center-of-mass}
\begin{matrix}
C_{x}=\frac{1}{N_a}\sum_{i=1}^{N_a} x_{i}w_{i} \\ 
C_{y}=\frac{1}{N_a}\sum_{i=1}^{N_a} y_{i}w_{i} \\ 
C_{z}=\frac{1}{N_a}\sum_{i=1}^{N_a} z_{i}w_{i}.
\end{matrix}
\end{equation}
Projection $t_{i}$ of an atom $i$ on the principal axis is given by:
\begin{equation}
\label{eq-projection-ti}
t_{i}=a_{x}(x_{i}-C_{x})+a_{y}(y_{i}-C_{y})+a_{z}(z_{i}-C_{z}).
\end{equation}
We then define a function $F$ as:
\begin{equation}
\label{eq-function-F}
F=\sum_{i}w_{i}t_{i}^{2}-\lambda (|a|^{2}-1).
\end{equation}
The function $F$ is now the target for maximization to find the possible longest principal axis onto which all atoms can be projected. 
The derivatives of $F$ over the coordinate axes:
\begin{equation}
\frac{\partial F}{\partial a_{x}}=\frac{\partial F}{\partial a_{y}}=\frac{\partial F}{\partial a_{z}}=0
\end{equation}
lead to an eigenvalue equation:
\begin{equation}
T
\begin{pmatrix}
a_{x}\\ 
a_{y}\\ 
a_{z}
\end{pmatrix}=\lambda \begin{pmatrix}
a_{x}\\ 
a_{y}\\ 
a_{z}
\end{pmatrix}.
\end{equation}
$T$ is the tensor of inertia about the center of mass with respect to the $xyz$ axes, written in a matrix form as:
\begin{equation}
\label{eq-inertia-tensor}
T=
\begin{pmatrix}
\sum_{i}w_{i}(Y_{i}^{2}+Z_{i}^{2}) &-\sum_{i}w_{i}X_{i}Y_{i}  &-\sum_{i}w_{i}X_{i}Z_{i} \\ 
-\sum_{i}w_{i}Y_{i}X_{i} &\sum_{i}w_{i}(X_{i}^{2}+Z_{i}^{2})  &-\sum_{i}w_{i}Y_{i}Z_{i} \\ 
-\sum_{i}w_{i}Z_{i}X_{i} &-\sum_{i}w_{i}Z_{i}Y_{i}  &\sum_{i}w_{i}(X_{i}^{2}+Y_{i}^{2}) 
\end{pmatrix},
\end{equation}
where 
\begin{equation}
\begin{matrix}
X_{i}=x_{i}-C_{x}\\ 
Y_{i}=y_{i}-C_{y}\\  
Z_{i}=z_{i}-C_{z}.\\ 
\end{matrix}
\end{equation}
Finally, the principal axis is found by solving the eigenvalue problem with the inertia tensor. After solving the eigenvalue problem, we obtain three eigenvalues and the corresponding eigenvectors. Among the three states, we choose the state that maximizes the function $F$ in Eq. (\ref{eq-function-F}). Also, the derivations imply that there are many ways to properly decompose a system by changing the definition of the function $F$. In our method, it is expressed as a maximization problem to find the correspondingly longest principal axis so that all atoms can be projected onto it. 
\begin{figure}[htbp]
\centering
\includegraphics[scale=0.8,trim=0cm 0cm 0cm 0cm]{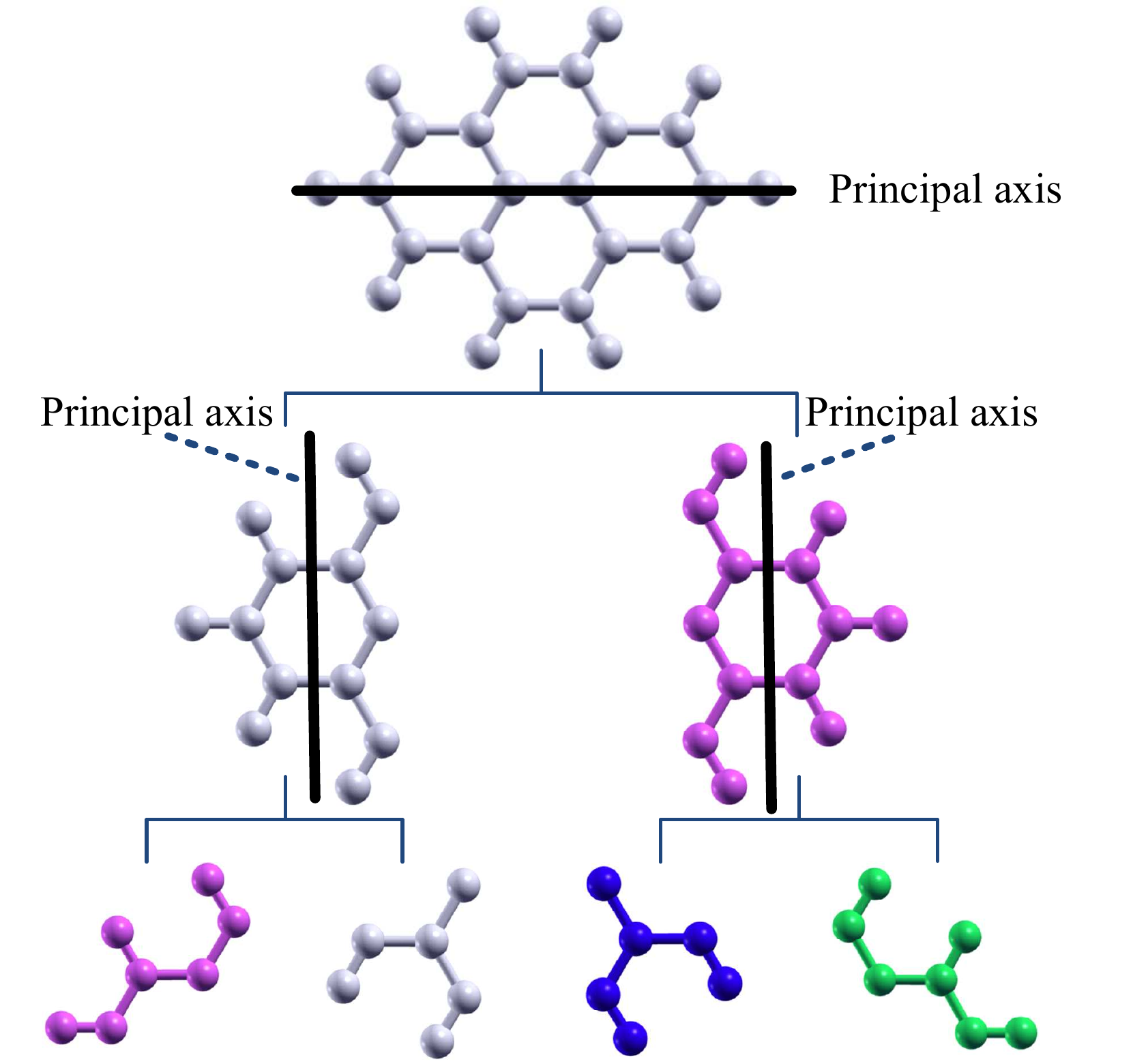}
\caption{Atom decomposition with the modified recursive bisection method and the principal axis.}
\label{fig-principle-axis}
\end{figure}

Figure \ref{fig-principle-axis} exemplifies the operation of our atom domain decomposition scheme based on the modified recursive bisection method and the principal axis. The binary tree is constructed with the root representing the original domain consisting of 26 atoms. The principal axis of the original domain is then found, based on which the domain is divided into two sub-domains with each having 13 atoms. The sub-domains are assigned to the two child nodes of the tree and the process of finding the principal axis and dividing the domain is repeated for each sub-domain. The domain on the left child node is divided into two sub-domains with one having 7 atoms and the other having 6 atoms, and so is the domain on the right child node. The decomposition is performed recursively until there are as many sub-domains as processes.

\subsubsection{Load balancing}
As different atoms may have different processing time depending on their properties, each atom is associated with a weight for load balancing. As a result, both the number of atoms and weights are considered in the decomposition method in an effort to distribute the amount of computation equally among the processes. The weight of an atom $i$ is defined as:
\begin{equation}
\label{eq-weight-def}
w_{i}=\frac{Etime_{i}}{Mtime},
\end{equation} 
where $Etime_{i}$ is the processing time of the atom $i$ in the previous MD step, and $Mtime$ is the longest processing time of all atoms. 

It should be noted that the weight definition has an important impact upon reordering the atoms. The weight defined in Eq. (\ref{eq-weight-def}) tends to position light atoms at around the center, while heavy atoms tend to move away from the center. As the bisection for dividing the domain is usually found near the center, the decomposition is indeed more precise when light atoms are also positioned around it. Therefore, our method based on the inertia tensor moment fulfils the requirement for load balancing and locality. 

\subsubsection{Algorithm for atom decomposition}
\label{the-algorithm}
\begin{figure}[htb]
\centering
\includegraphics[scale=0.8,trim=0cm 0cm 0cm 0cm]{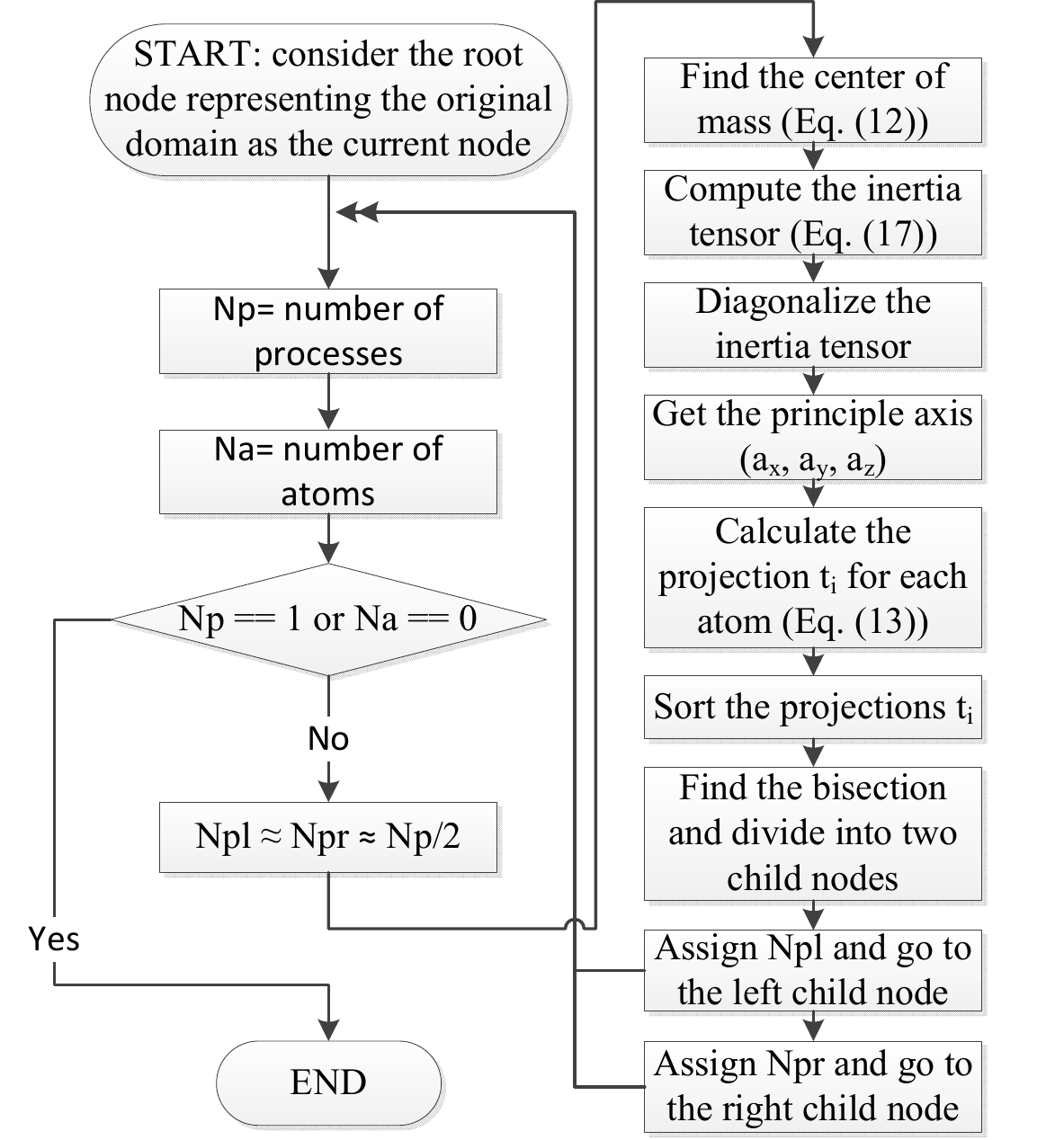}
\caption{The algorithm for decomposing atoms.}
\label{fig-algorithm-atoms}
\end{figure}

Figure \ref{fig-algorithm-atoms} outlines the algorithm for the atom decomposition method with a combination of the modified recursive bisection method and the inertia tensor. It starts with the root node of the binary tree representing the original domain as the current node. If the number of processes is 1 or the number of atoms is 0, the current node is the leaf node and the algorithm will stop. Otherwise, it will find the center of mass with weights of the current domain (Eq. (\ref{center-of-mass})), and compute the inertia tensor based on Eq. (\ref{eq-inertia-tensor}). After that, the inertia tensor is diagonalized to obtain the principle axis. Now the projection $t_{i}$ of each atom onto the principal axis can be determined using Eq. (\ref{eq-projection-ti}). They are then sorted to find the bisection for dividing the current node into two child nodes based on Eq. (\ref{eq-weight-def}) where the sum of weight is nearly equivalent to each other.  
%, each with approximately the same numbers of atoms and processes. 
The algorithm repeats with each child node until reaching the leaf node, where there is only one assigned process or no atom available.    

\subsection{Three-dimensional grid decomposition method}
\label{grid-decomposition}
As the regular mesh technique is used in our implementation \citep{PhysRevB.72.045121}, after the atoms have been distributed to the processes, the grid points belonging to their basis functions must also be allocated to the corresponding ones. Once the density matrix $\rho$ has been calculated by Eq. (\ref{density-matrix}), the charge density can be found by Eq. (\ref{charge-density}) with the basis functions using regular mesh. As the Hartree potential needs to be calculated by FFT using the whole system, the allocation of the grids to the processes should be carried out effectively so that the communication amount among them is as small as possible. 

\subsubsection{Requirements}
In parallel with the atom decomposition method, a 3D domain decomposition method for grids is also developed with the following requirements. 
\begin{itemize}
\item
{\bf Approximately the same number of grid points}: Similar to the case of atom decomposition, processes should be assigned nearly the same number of grid points so that they can perform approximately the same computational amount.  
\item
{\bf Almost linear scaling in memory usage}: Memory usage is always a source of concern in DFT-based calculations, as a large amount of memory is demanded for storing a huge number of grid points. As such, the amount of memory required for serial calculations should be only linear to the system size, and should remain a constant for parallel calculations as a function of the numbers of atoms and processes when the same number of atoms is allocated to each process.        
\item
{\bf Locality}: Grid points within the sphere of a basis function (Eq. (\ref{basis-function})) belonging to an atom that is allocated to a process should be assigned to the same process with the atom. The locality should also be held when distributing the grids to the processes. This will have two-fold benefits: minimize the amount of communication and the amount of memory necessary for storing data on the grid points overlapping with those of the basis functions for all atoms in one process.      
\end{itemize}

\subsubsection{Data structures for MPI parallelization}
\label{data-structure}
\begin{figure}[htb]
\centering
\includegraphics[scale=0.8,trim=0cm 0cm 0cm 0cm]{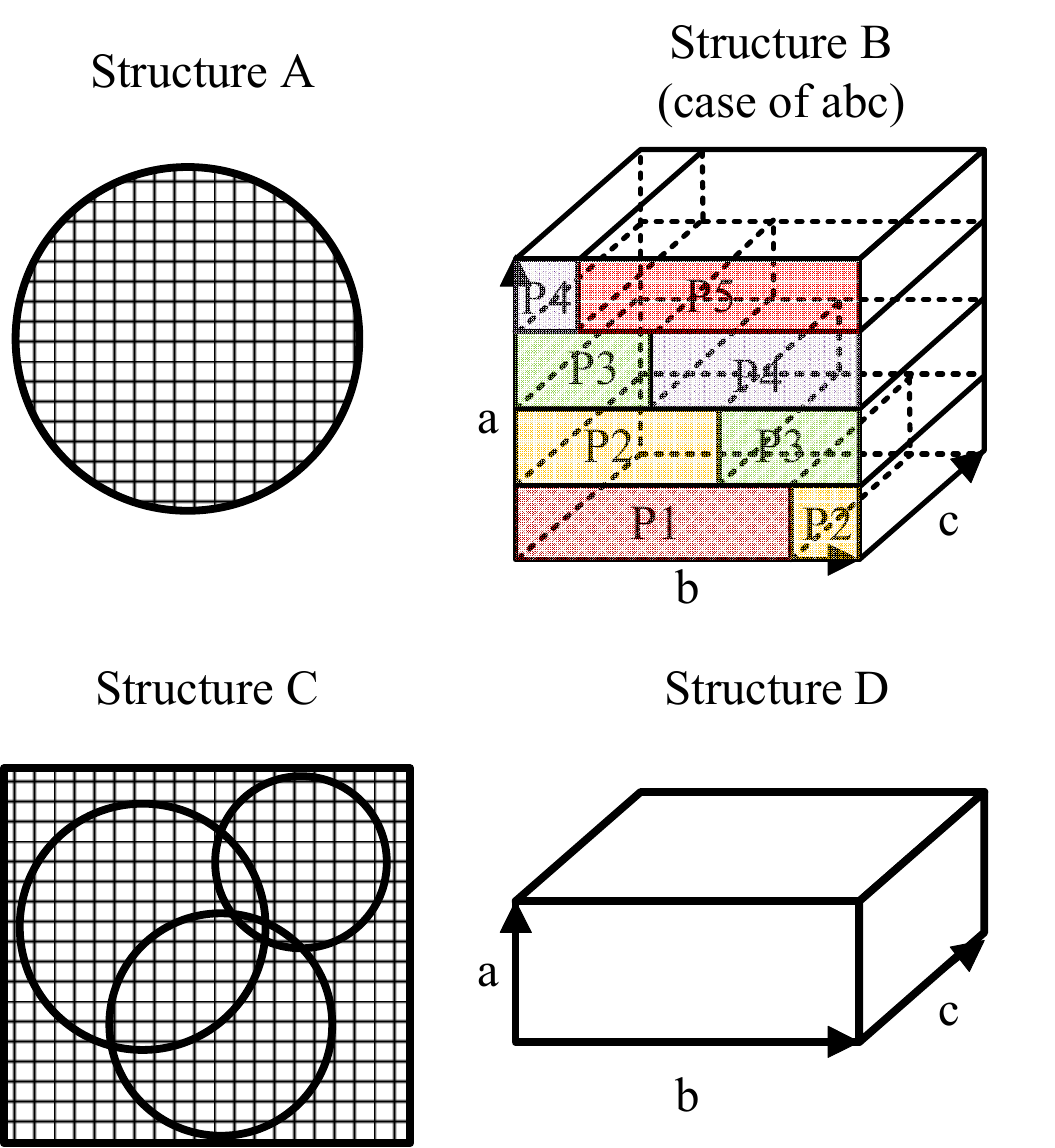}
\caption{Data structures for MPI parallelization.}
\label{fig-grid-structure}
\end{figure}

To comply with the requirements, here we define four data structures to make the data locality consistent with that of the clustered atoms for minimizing the amount of communication between the processes, which are shown in Fig. \ref{fig-grid-structure}.

\begin{itemize}
\item
{\bf Structure A}: This structure stores all the grid points inside the sphere of a basis function. 
\item
{\bf Structure B}: This structure stores all the grid points allocated to a specific process. Here we propose a two-dimensional decomposition method for parallel 3D FFT, where the distribution of the grid points is carried out in a two-dimensional grid defined by the first two dimensions. Therefore, the structure B may exist in different forms by changing the order of distribution. For example, in case of $\textit{\textbf{abc}}$ distribution, the grid points on the $\textit{\textbf{ab}}$-plane with the $\textit{\textbf{c}}$ axis are divided over the processes in ascending order of their $\textit{\textbf{a}}$ and $\textit{\textbf{b}}$ coordinates so that each process has approximately the same number of grid points on the $\textit{\textbf{ab}}$-plane. The grid points extending along the $\textit{\textbf{c}}$ direction that have the same $\textit{\textbf{a}}$ and $\textit{\textbf{b}}$ coordinates are also assigned to the same process. Assume that the numbers of grids of the \textit{\textbf{a}}-, \textit{\textbf{b}}-, and \textit{\textbf{c}}-axes are $N_{\mathrm{a}}$, $N_{\mathrm{b}}$, and $N_{\mathrm{c}}$, respectively, the number of processes is $N_p$, and $myid$ is the process's ID. 
%Let $N_{\mathrm{m}}=\left \lceil \frac{N_{\mathrm{a}}N_{\mathrm{b}}}{N_{\mathrm{p}}} \right \rceil$ be the number of grid points on the $\textit{\textbf{ab}}$-plane to be distributed to each process, except for the last process that may be allocated one grid point fewer than the others. 
In our method, the process with $myid$ will be assigned the grid points from the coordinate of $(a_s, b_s, 0)$, where $a_{s}N_{\mathrm{b}}+b_{s}=\left \lfloor \frac{myid \times N_{\mathrm{a}}N_{\mathrm{b}}+N_{\mathrm{p}}-1}{N_{\mathrm{p}}} \right \rfloor$, to the coordinate of $(a_e, b_e, N_{\mathrm{c}}-1)$, where $a_{e} N_{\mathrm{b}}+b_{e}=\left \lfloor \frac{(myid+1) \times N_{\mathrm{a}}N_{\mathrm{b}}+N_{\mathrm{p}}-1}{N_{\mathrm{p}}} -1\right \rfloor$, in ascending order of their $\textit{\textbf{a}}$, $\textit{\textbf{b}}$, and $\textit{\textbf{c}}$ coordinates. As a result, our decomposition method can be viewed as a hybrid method between one-dimensional decomposition and two-dimensional decomposition, since it is the same as the former for up to a certain number of processes, and is gradually akin to the latter for larger numbers of processes. Detailed analysis on the communication amount will be given later in \ref{MPI-communication-analysis}. 
\item
{\bf Structure C}: The structure stores all the grid points inside a parallelepipedon defined by the basis functions of all atoms allocated to one process, paying attention to the cutoff radius of the basis functions.    
\item
{\bf Structure D}: This structure can be seen as an extension of the structure B(\textit{\textbf{abc}}) because it is created by adding some buffer regions around the structure B. Strictly speaking, this structure is only necessary for GGA \cite{perdew1996generalized} as it requires neighboring grid points for calculating the gradient of the charge density by a finite difference method. Meanwhile, LDA \cite{perdew1981self} can be performed with the structure B without having to refer to the structure D. 
\end{itemize}

The introduction of the data structures makes it more straightforward in the calculations. For example, one can easily find a specific basis function by referring to the structure A.
%, or the grid points belonging to all basis functions in a particular processes using the structure C. 
As can be seen subsequently, the structure B is related to a two-dimensional decomposition method for FFT that is more efficient than previously proposed methods in terms of communication amount. 

\subsubsection{MPI communication analysis in the domain decomposition method for 3D FFT}
\label{MPI-communication-analysis}

\begin{figure}[htbp]
\begin{center}
\subfigure[Our method.]{\label{fig-MPI-B1}
\includegraphics[scale=0.75]{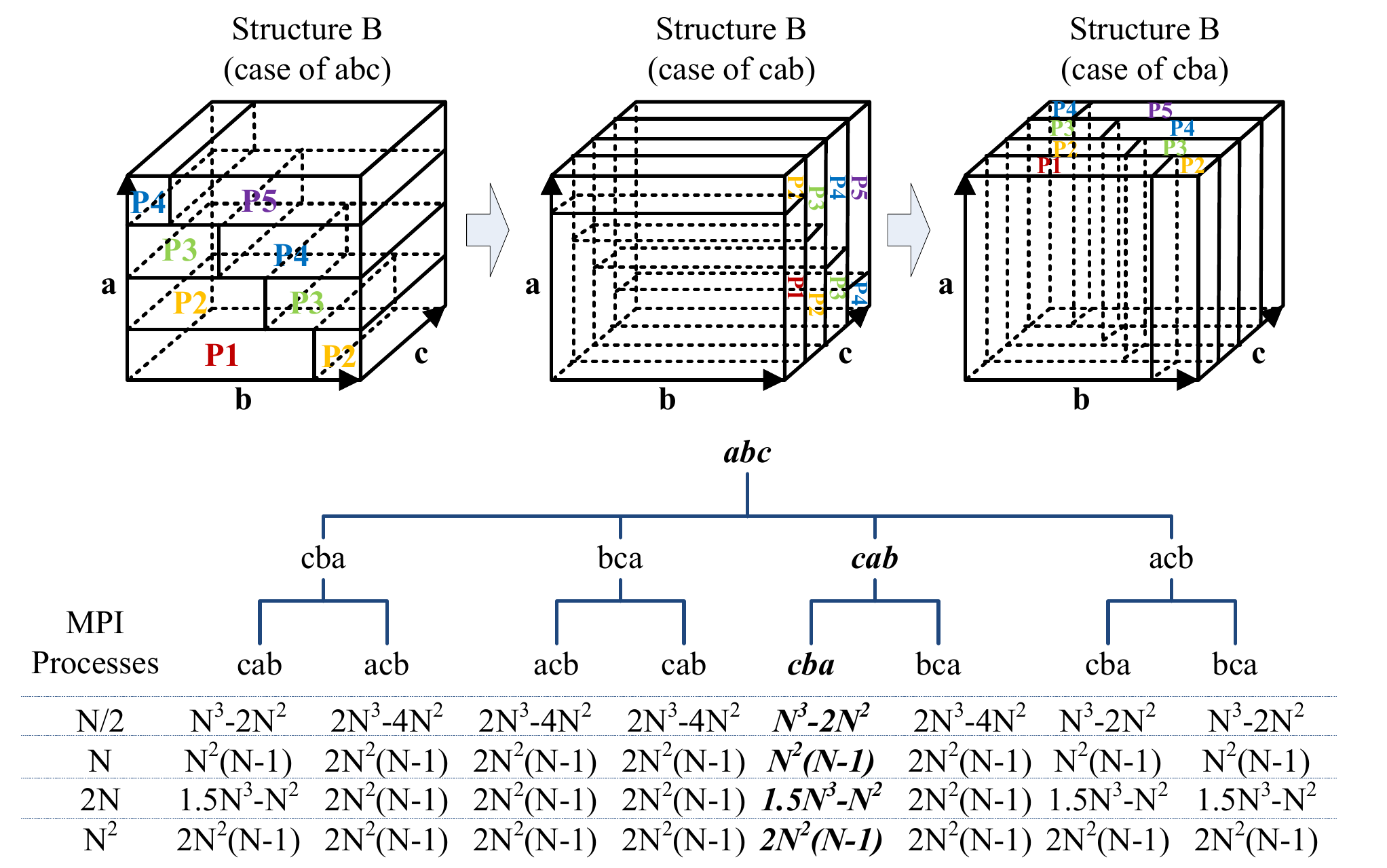}}
\subfigure[One- and two-dimensional methods.]{\label{fig-1D-2D}
\includegraphics[scale=1]{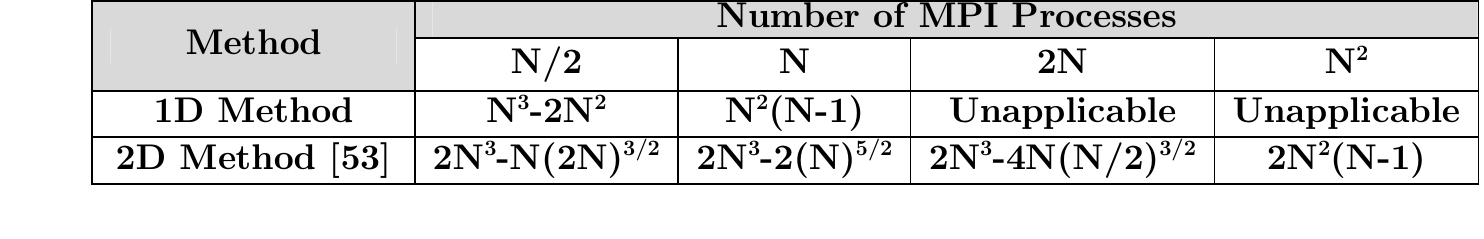}} \\
\end{center}
\caption{MPI communication analysis in the parallelization of 3D FFT for (a) our method and (b) other one-dimensional method and two-dimensional method \cite{takahashi2010implementation} with $N \times N \times N$ grid points.}
\label{fig-MPI-B}
\end{figure}

%\begin{figure}[htb]
%\centering
%\includegraphics[scale=0.75,trim=0cm 0cm 0cm 0cm]{fig-MPI-B.pdf}
%\caption{MPI communication analysis in structure B with $N \times N \times N$ grid points.}
%\label{fig-MPI-B}
%\end{figure}

%Structure B appears to be the most important structure that incurs communications among the processes, especially when performing FFT. 
In our two-dimensional decomposition method for parallel 3D FFT, as different distributions of the structure B are involved in the calculations in different orders, the order of reference may have an impact on communication amount. Figure \ref{fig-MPI-B} shows an analysis on the amount of communication to transpose the structure B from one to another depending on the number of processes, in comparison with a traditional one-dimensional decomposition method and a two-dimensional decomposition method \cite{takahashi2010implementation}. Interestingly, the amount of communication is grouped into only two patterns of communication in our method, and there is one pattern actually better than the other. In general, we can choose any of the four orders leading to the pattern with the smaller amount of communication, for instance $\textit{\textbf{abc}}$ $\rightarrow$ $\textit{\textbf{cab}}$ $\rightarrow$ $\textit{\textbf{cba}}$, or $\textit{\textbf{abc}}$ $\rightarrow$ $\textit{\textbf{cba}}$ $\rightarrow$ $\textit{\textbf{cab}}$. However, we choose the order of $\textit{\textbf{abc}}$ $\rightarrow$ $\textit{\textbf{cab}}$ $\rightarrow$ $\textit{\textbf{cba}}$ to keep consistency with our current implementation of other functionalities such as non-equilibrium Green's function (NEGF) method.  
%However, considering the fact that we should make the $\textit{\textbf{a}}$ axis fit to our current implementation, we choose the order of $\textit{\textbf{abc}}$ $\rightarrow$ $\textit{\textbf{cab}}$ $\rightarrow$ $\textit{\textbf{cba}}$.  

\begin{figure}[htb]
\centering
\includegraphics[scale=1,trim=0cm 0cm 0cm 0cm]{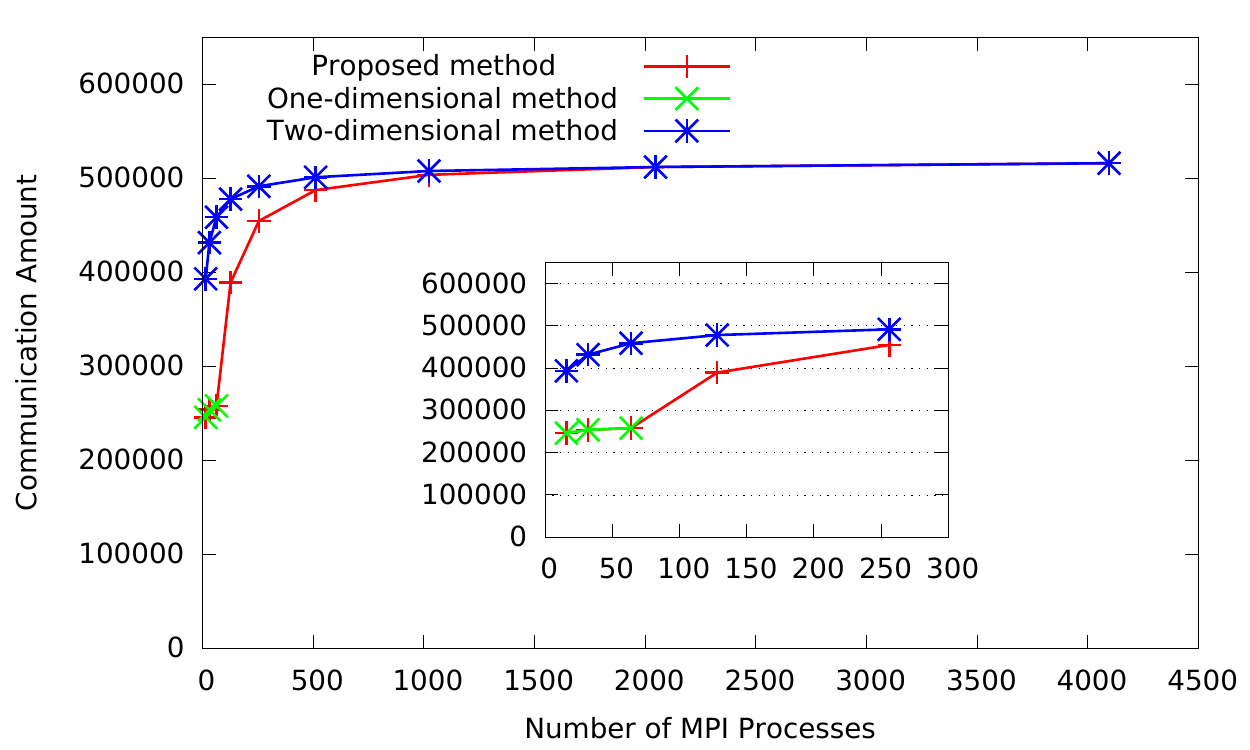}
\caption{Comparison of communication amount in the parallelization of 3D FFT in case of $64 \times 64 \times 64$ grid points ($N=64$). The inset enlarges the left part of the main graph.}
\label{fig-B-compare}
\end{figure}

Figure \ref{fig-B-compare} shows a comparison among our proposed method and the two other methods in terms of communication amount in case of $64 \times 64 \times 64$ grid points. Certainly, similar results can be produced for smaller and larger numbers of grid points. In this case, the one-dimensional method is able to work only along one dimension and is limited to 64 processes, while the two-dimensional method always decomposes the domain in two dimensions, even for fewer than 64 processes. In contrast, our method offers a combination of these methods, as it partitions only along one dimension when the number of processes is up to 64 on condition that it is a divisor of 64, and decomposes in two dimensions while still starting from one dimension for larger process numbers. In other words, our method is based on a row-wise decomposition, as against a square decomposition approach taken in \cite{takahashi2010implementation}. In doing so, most of communication amount is incurred when transferring from $\textit{\textbf{abc}}$ to $\textit{\textbf{cab}}$, and a majority of data are reused when transferring from $\textit{\textbf{cab}}$ to $\textit{\textbf{cba}}$ with just a small amount of communication. As a result, up to 64 processes, our method works in the same fashion as the one-dimensional method provided that 64 is a multiple of the number of processes, and is about 60.0\% to 77.8\% better than the two-dimensional method with $64 \times 64 \times 64$ grid points. Beyond this point, the one-dimensional method is no longer applicable, while the two other methods can operate until reaching the limit of $64 \times 64 = 4096$ processes. The difference in performance gradually decreases, however, and eventually becomes 0 from 2048 processes.
%, when our method and the two-dimensional method share the same decomposition implementation. 

%Note on the communication amounts
%Proposed method
%N/4: N^3-4N^2; N/2: N^3-2N^2; N: N^3-N^2; 2N: 1.5N^3-N^2; 4N: 7/4*N^3-N^2; 8N: 15/8*N^3-N^2; 16N: 31/16*N^3-N^2; 32N: 63/32*N^3-N^2;  

%Takahashi method
%N/4: 2N^3-N/2*(4N)^(3/2); N/2: 2N^3-N*(2N)^(3/2); N: 2N^3-2*(N)^(5/2); 2N: 2N^3-4N*(N/2)^(3/2); 4N: 2N^3-8N*(N/4)^(3/2); 8N: 2N^3-16N*(N/8)^(3/2); 16N: 2N^3-32N*(N/16)^(3/2); 32N: 2N^3-64N*(N/32)^(3/2);  

\subsubsection{Data structures and calculation flow}
\begin{figure}[htb]
\begin{center}
\subfigure[Data structures of grid decomposition.]{\label{fig-data-structure}
\includegraphics[scale=0.8]{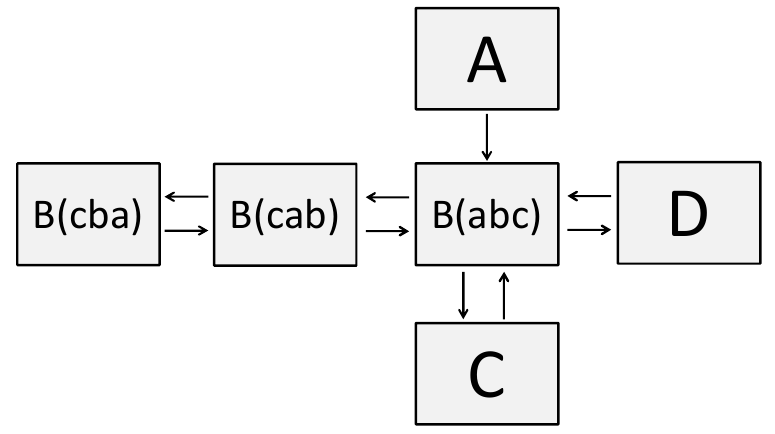}}
\subfigure[Calculation flow.]{\label{fig-calculation-flow}
\includegraphics[scale=0.8]{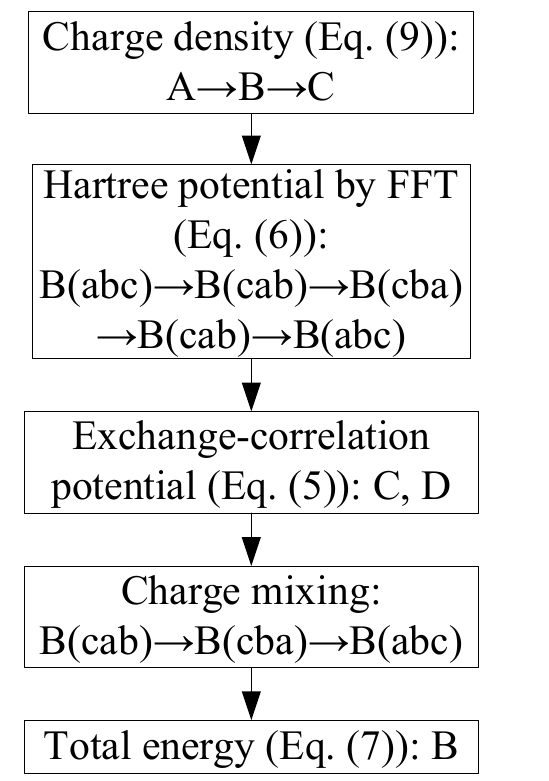}} \\
\end{center}
\caption{(a) Data structures and (b) calculation flow.}
\label{fig-data-calculation}
\end{figure}

The relationship between the data structures and the calculation flow is depicted in Fig. \ref{fig-data-calculation}. The calculation of charge density (Eq. (\ref{charge-density})) is performed with the structures A, B, and C in order of appearance. The Hartree potential (Eq. (\ref{Hartree-potential})) is calculated by solving the Poisson equation using FFT for the whole system with different distributions of the structure B in the order of $\textit{\textbf{abc}}$ $\rightarrow$ $\textit{\textbf{cab}}$ $\rightarrow$ $\textit{\textbf{cba}}$ $\rightarrow$ $\textit{\textbf{cab}}$ $\rightarrow$ $\textit{\textbf{abc}}$. The exchange-correlation potential (the last term in Eq. (\ref{eq-KS-potential})) is determined from either the structure C or the structure D depending on the approximation method in use. The structure B is also utilized for performing the charge mixing ($\textit{\textbf{cab}}$ $\rightarrow$ $\textit{\textbf{cba}}$ $\rightarrow$ $\textit{\textbf{abc}}$), and for calculating the total energy (Eq. (\ref{total-energy})). It should be noted that communication is required when the calculations move from one structure to another. 

%\begin{figure}[htb]
%\centering
%\includegraphics[scale=0.8,trim=0cm 0cm 0cm 0cm]{fig-data-calculation.pdf}
%\caption{Data structures and the calculation flow.}
%\label{fig-data-calculation}
%\end{figure}

%\subsubsection{Limitations}
%\subsection{Memory management}

\section{Implementation}
\label{Implementation}
\begin{figure}[htb]
\centering
\includegraphics[scale=0.7,trim=0cm 0cm 0cm 0cm]{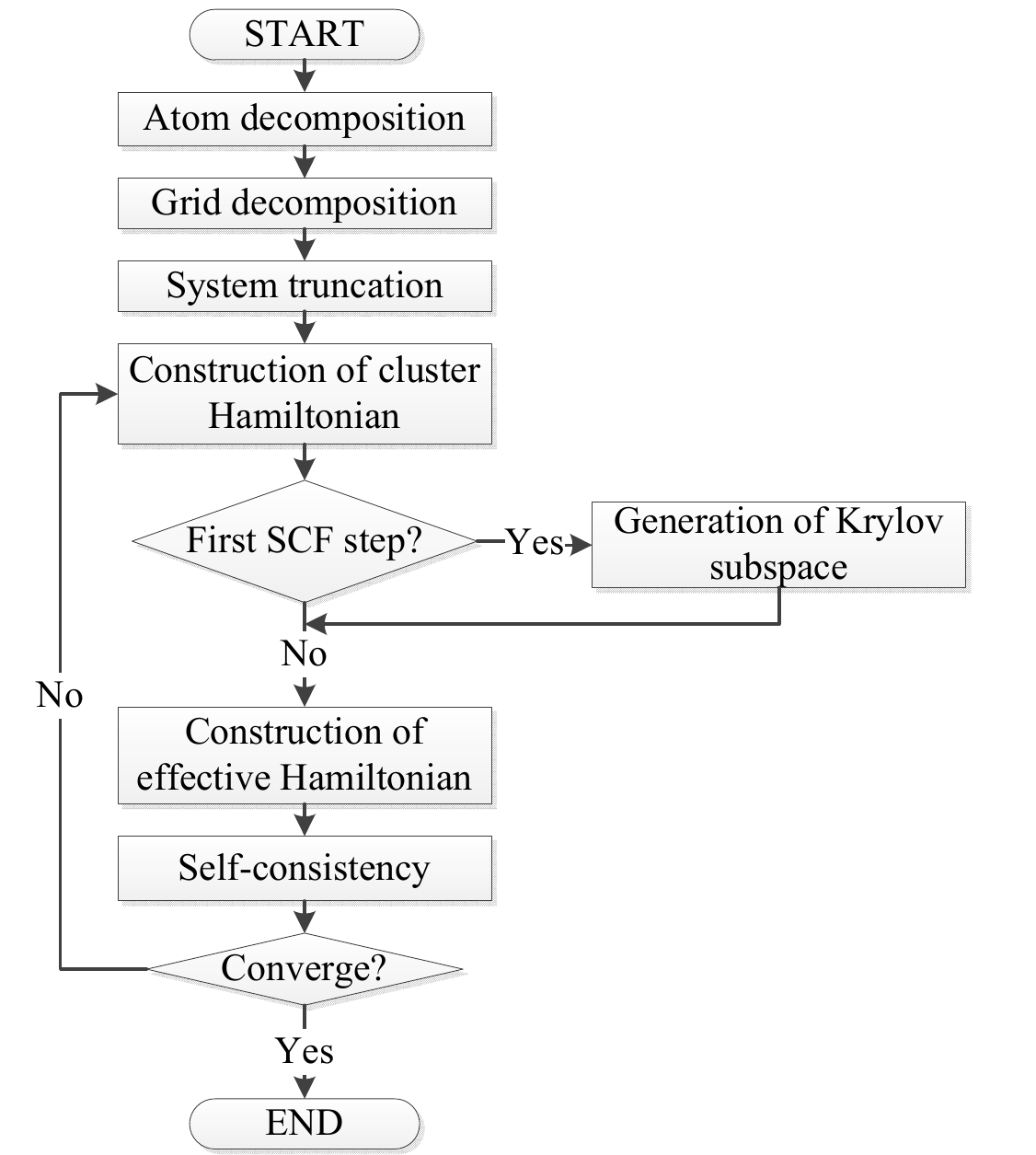}
\caption{Computational flow.}
\label{fig-implementation-chart}
\end{figure}

Figure \ref{fig-implementation-chart} illustrates the implementation flowchart in terms of the computational flow that consists of three key steps: atom decomposition, grid decomposition, and $O(N)$ calculation. The first two steps are common, and applicable to other schemes such as conventional DFT calculations, NEGF method, etc., while the last step is specially for our linear scaling Krylov subspace method. 

\begin{enumerate}[Step 1]
\item
{\bf Atom decomposition.}
The atoms are allocated to the processes by the proposed atom decomposition method based on a combination of the modified recursive bisection method and inertia tensor. The algorithm describing its operation is detailed in section \ref{the-algorithm}. 
\item
{\bf Grid decomposition.}
The grid points are decomposed using our grid decomposition method. It constructs the structures A, B, C, and D based on their definition. The data structure for transferring $\rho_i(\textbf{\textit{r}})$ from the structure A to the structure B when the charge density $\rho(\textbf{\textit{r}})$ is calculated in the structure B using $\rho_i(\textbf{\textit{r}})$ is constructed. Similarly, the data structure for transferring $\rho(\textbf{\textit{r}})$ from the structure B to the structure C when $\rho(\textbf{\textit{r}})$ is constructed in C using $\rho(\textbf{\textit{r}})$ stored in B, and the data structure for transferring $\rho(\textbf{\textit{r}})$ from the structure B to the structure D for the case that $\rho(\textbf{\textit{r}})$ is constructed in D using $\rho(\textbf{\textit{r}})$ stored in B are constructed. In order to calculate the Hartree potential by FFT with different distributions of the structure B, the data structures for MPI communications in FFT from $\textit{\textbf{abc}}$ to $\textit{\textbf{cab}}$ and from $\textit{\textbf{cab}}$ to $\textit{\textbf{cba}}$ are also built. It is worth mentioning that we employ non-blocking MPI routines MPI\_Isend() and MPI\_Irecv() for point-to-point communications between pairs of processes to cut the number of communication steps between the processes and hence, reducing their waiting time. 
%exploit the computation and communication overlap whenever possible. 
This implementation is definitely much more efficient than posting pairs of blocking MPI\_Send() and MPI\_Recv().  
\item
{\bf $O(N)$ calculation.}
The linear scaling Krylov subspace method is implemented in this step. Although the atoms have already been allocated nearly equally to the MPI processes in Step 1, here we further decompose the allotted atoms to the processing cores in each MPI process using OpenMP thread parallelization in the hybrid MPI/OpenMP implementation. 
\begin{enumerate}[Step 3a]
\item
{\bf System truncation:}
%Physical and logical truncation schemes \cite{ozaki2000block} are employed to divide the large system into smaller truncated clusters. The physical scheme collects all the neighboring sites within a sphere with a certain cutoff radius, while the logical scheme defines a truncated cluster containing all the neighboring sites that can be reached in a certain number of hops. The logical scheme is more stable and effective in preventing sudden energy jumps in MD simulations from occurring, leading to higher accuracy. 
Physical truncation scheme \cite{ozaki2000block} is employed to divide the large system into smaller truncated clusters. It collects all the neighboring sites within a sphere with a certain cutoff radius to construct the corresponding clusters.    
\item
{\bf Construction of cluster Hamiltonian:}
The Hamiltonian of each truncated cluster is constructed in the conventional fashion \citep{PhysRevB.72.045121}. The matrix elements in each site can be calculated independently in parallel, and the cluster Hamiltonian is built by gathering all the elements from the processes through communications. Also, the Hartree potential is determined from all the contributions from the truncated clusters to ensure high accuracy.  
\item
{\bf Generation of Krylov subspace:}
For numerical robustness, the Krylov subspace of each truncated cluster is generated only at the first SCF step, and is re-utilized in subsequent steps \citep{ozaki2006n}. The Krylov subspace generation is independent and performed in parallel. 
\item
{\bf Construction of effective Hamiltonian:}
The effective Hamiltonian is constructed by determining the short and long-range contributions, which can be performed in parallel on each process. For truncated clusters with a large buffer region, the long-range contribution is calculated only at the first SCF step and re-used in subsequent steps. 
\item
{\bf Self-consistency:}
The standard eigenvalue problem with the effective Hamiltonian is solved, and the eigenvectors are obtained by performing the back transformation to calculate the charge density. After that, a common chemical potential that conserves the total number of electrons for all the truncated clusters is determined by a bisection method, where MPI communication is performed using MPI\_Allreduce \citep{ozaki2006n}. 
%where the processes exchange the number of electrons in each truncated cluster \citep{ozaki2006n}. 
\end{enumerate}
\end{enumerate}

\section{Benchmark results}
\label{Benchmark-results}
\subsection{Software and system configuration}
\subsubsection{OpenMX}
\begin{table}
\centering
\caption{Software configuration.}
\includegraphics[scale=1,trim=2cm 0.5cm 0cm 0cm]{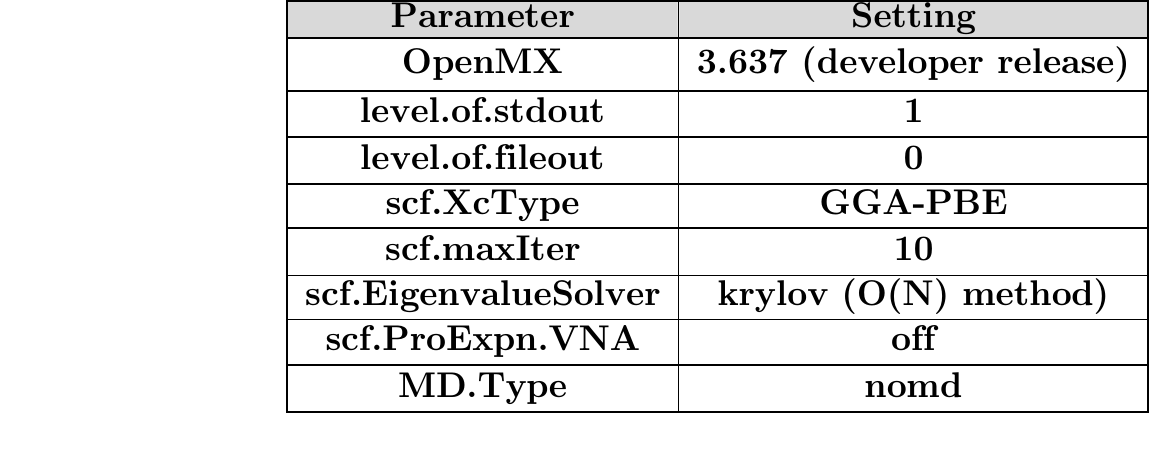}
\label{tab-soft-config}
\end{table}
All the calculations are performed using OpenMX \citep{openmx}, which is an open source parallel software package developed based on DFT, norm-conserving pseudopotentials, and pseudo-atomic localized basis functions. In particular, we use OpenMX version 3.6, release 37 (developer release) with a focus on the linear scaling Krylov subspace method. The number of SCF iterations is set at 10 for each calculation with no geometry optimization. In addition, GGA is employed to calculate the exchange-correlation potential, a required energy cutoff of 150 Ry is specified in the real space grid techniques for numerical integrations and solving the Poisson equation using FFT, while the cutoff energy is determined by the unit cell size to minimize the cutoff energy difference in the $a-$, $b-$, and $c-$axes, and the electronic temperature is set at 300 K for counting the electron number. With regard to the Krylov subspace method, the radius of a sphere centered on each atom is fixed at 6 Ang, and the dimension of the Krylov subspace in each truncated cluster is set at 400. As a result, the number of atoms in the truncated cluster is 159 with the diamond structure. With these settings in our method, we confirm that the absolute error in the total energy is 0.00072 Hartree/atom in self-consistent calculations with the diamond structure, in comparison to the $k$-space conventional method performed with a large number of $k$ points. Table \ref{tab-soft-config} summarizes some important parameters.      

\subsubsection{System configuration}
\begin{table}
\centering
\caption{System configuration.}
\includegraphics[scale=1,trim=2cm 0.5cm 0cm 0cm]{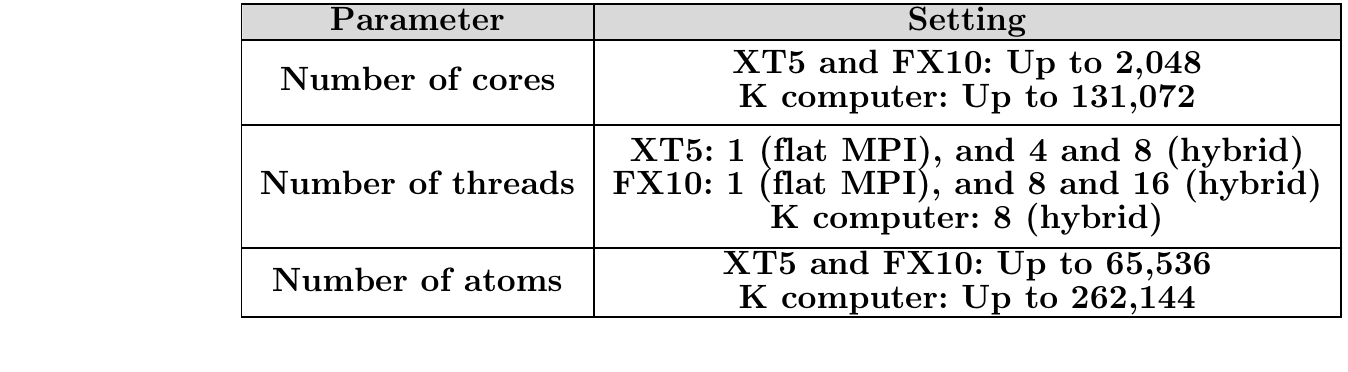}
\label{tab-sys-config}
\end{table}
We perform the series of benchmark calculations on three different machines: Cray XT5, Fujitsu FX10, and the K computer. Each XT5 node is composed of two quad-core AMD Opteron 2.4GHz with 16GB memory, coupled with Cray SeaStar2+ interconnect, while each FX10 node has one 1.848GHz 16-core SPARC64 IXfx processor, 32GB memory, and Tofu interconnect. Each K computer node contains one 2.0GHz 8-core SPARC64 VIIIfx processor, 16GB memory, and the same Tofu interconnect as FX10. We use up to 2,048 cores on the first two machines and 131,072 cores on the K computer. On the other hand, as OpenMX is capable of executing in both flat MPI and hybrid MPI/OpenMP modes, several combinations of the number of processes and threads are applied to particular system architecture. Specially, the calculations are carried out with 1 thread (flat MPI), 4 and 8 threads (hybrid) on XT5, 1, 8, and 16 threads on FX10, and 8 threads on the K computer. The system configuration is shown in Table \ref{tab-sys-config}. 

\subsubsection{Benchmark system}
We mainly employ the diamond structure, which is well-suited to benchmark calculations due to its uniformity, with the basis function of C5.0-$s2p2$, where C represents the atomic symbol, 5.0 is the cutoff radius (Bohr) in the generation by the confinement method, and $s2p2$ indicates that two radial functions are used to expand the basis functions for $s$- and $p$- orbitals, respectively. The numbers of atoms used on XT5 and FX10 are 2,048, 4,096, 8,192, 16,384, 32,768, and 65,536 although we confirmed that systems of up to 131,072 atoms are runnable on XT5. For the K computer, we test with systems of up to 262,144 atoms. The number of atoms is also shown in Table \ref{tab-sys-config}. Furthermore, in order to perform test calculations with a structure different from the uniform diamond structure, we choose the long deoxyribonucleic acid (DNA) structure consisting of cytosines and guanines with 2,600, 13,000, and 26,000 atoms, and the basis functions of H7.0-$s2p1$, C7.0-$s2p2d1$, N7.0-$s2p2d1$, O7.0-$s2p2d1$, and P9.0-$s2p2d1$.  

\subsection{Results}
To minimize the impact of system performance variations, each calculation is executed five times, except for the K computer, and the average values are taken as the results of the calculation. The following subsections present the calculation results, divided into a demonstration of the 3D atom decomposition method, linear scaling property by the Krylov subspace method, strong and weak scaling properties with the diamond structure, strong scaling property with the DNA structure, and strong scaling property of sub-routines, followed by an analysis on memory usage. 

\subsubsection{3D atom decomposition}
\begin{figure}[htbp]
\begin{center}
\subfigure[Diamond: Top view.]{\label{fig:atom-top}
\includegraphics[scale=0.5,trim=0cm 0cm 0cm 0cm]{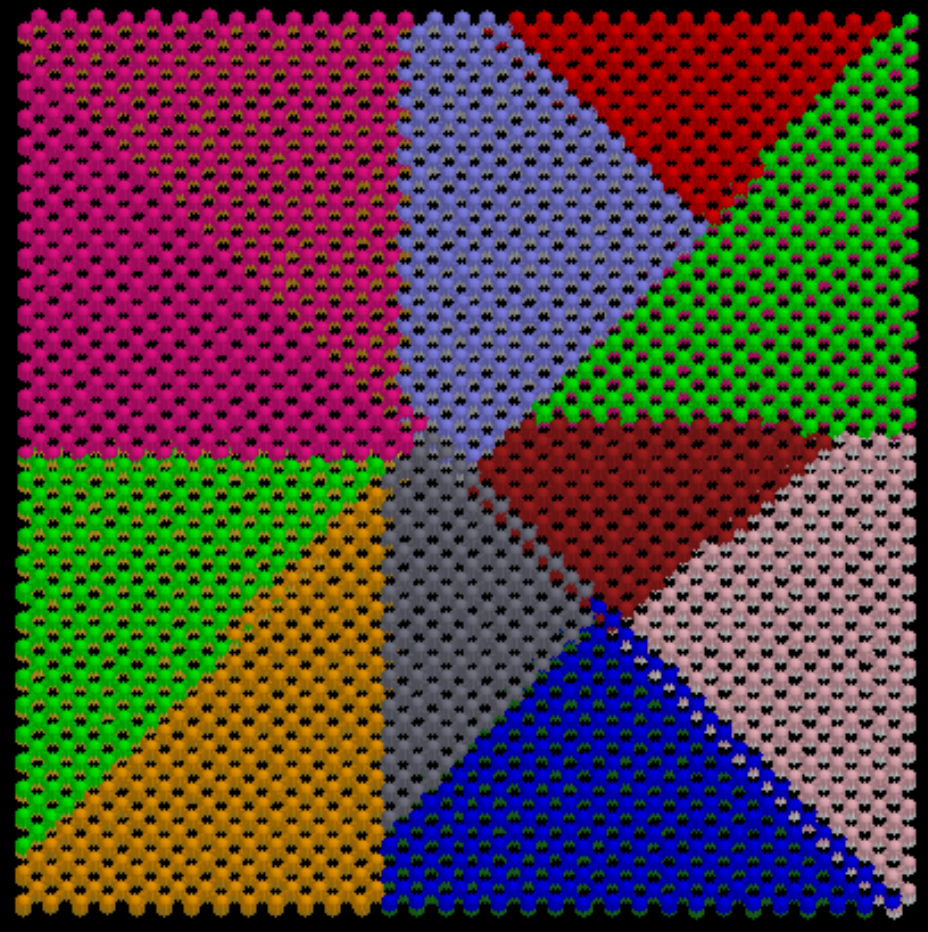}}
\subfigure[Diamond: Side view.]{\label{fig:atom-side}
\includegraphics[scale=0.5,trim=0cm 0cm 0cm 0cm]{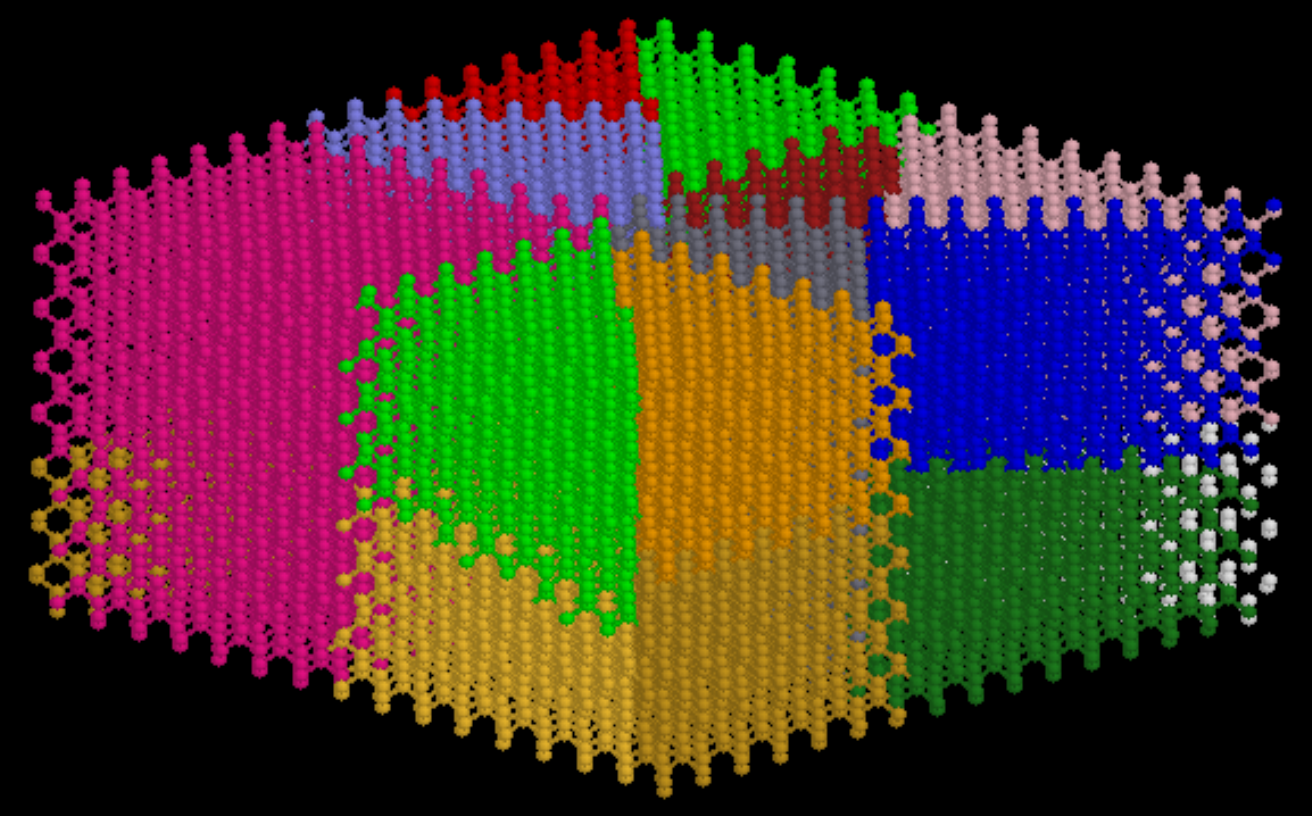}} 
\subfigure[Multiply-connected carbon nanotube: 8 MPI processes.]{\label{fig:CNT-a}
\includegraphics[scale=0.5,trim=0cm 0cm 0cm 0cm]{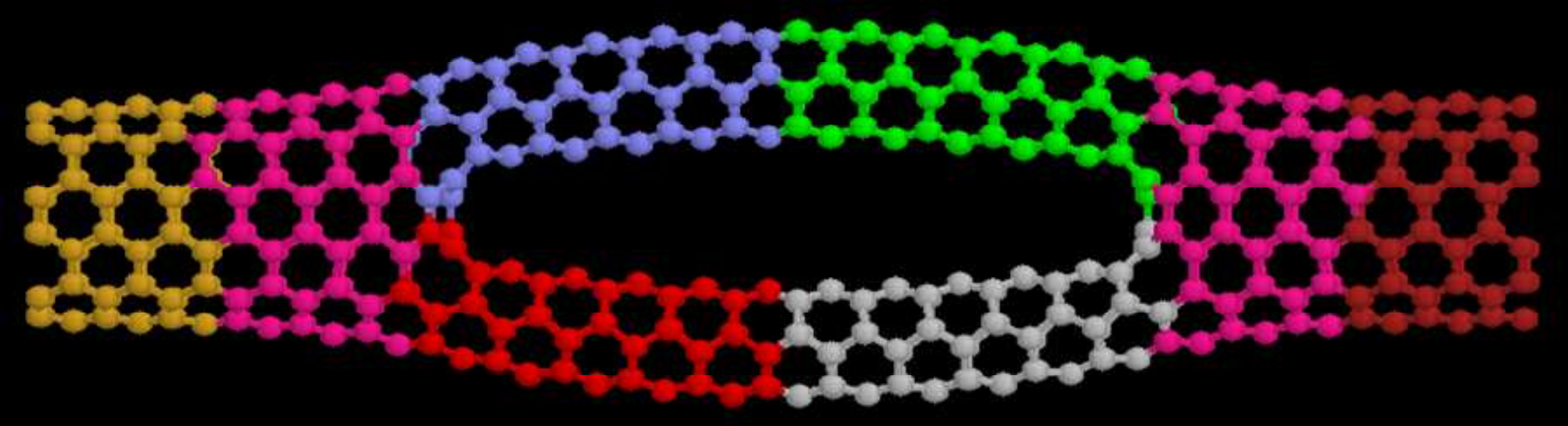}} 
\subfigure[Multiply-connected carbon nanotube: 16 MPI processes.]{\label{fig:CNT-b}
\includegraphics[scale=0.5,trim=0cm 0cm 0cm 0cm]{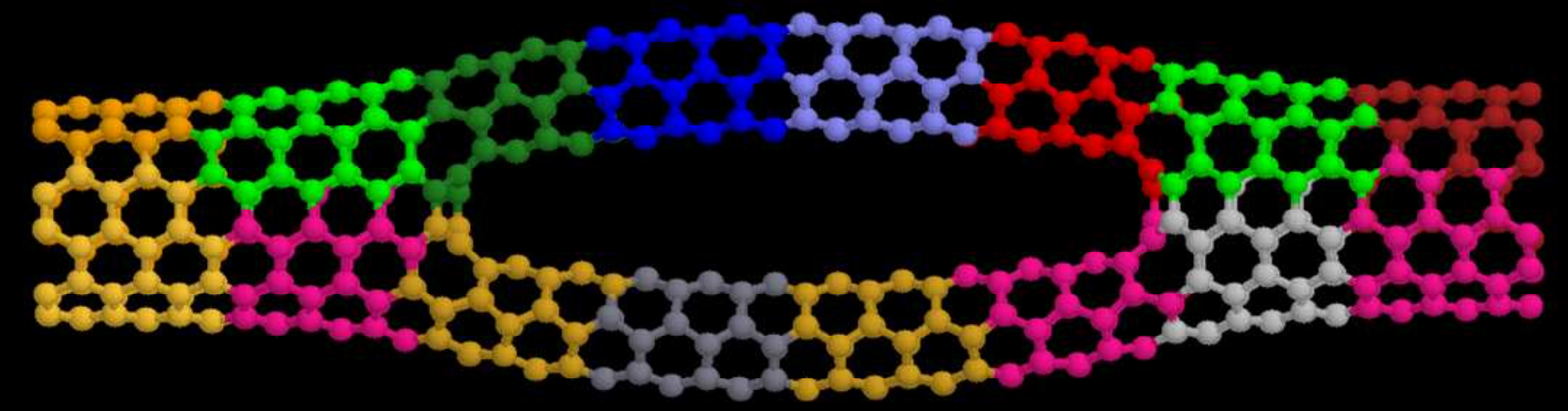}} 
\end{center}
\caption{3D atom decomposition examples: 16,384 diamond atoms and 19 MPI processes with (a) top view and (b) side view, and multiply-connected carbon nanotubes consisting of 564 carbon atoms with (c) 8 MPI processes and (d) 16 MPI processes. Atoms of the same color are allocated to the same MPI process.}
\label{fig-3d-atom}
\end{figure}
Figure \ref{fig-3d-atom} demonstrates the operation of our 3D atom decomposition in the cases of 16,384 diamond atoms and 19 MPI processes, and multiply-connected carbon nanotubes consisting of 564 carbon atoms with 8 and 16 MPI processes, where atoms of the same color are allocated to the same MPI process. In the diamond case, even though 19 is a prime number meant to pose challenges to general decomposition schemes, our method proves to be able to work efficiently, and atoms are practically localized to the process holding them.  Similar atom locality conservation is also obtained with the multiply-connected carbon nanotubes, where nearby atoms are distributed to the same process. This is of great importance to reduce communication amount and memory usage for performance enhancement. Although our examples are limited to up to 19 processes, it is guaranteed that the systems are well divided for many processes, since the decomposition is recursively applied to sub-systems.

\subsubsection{Linear scaling property}
\begin{figure}[htbp]
\begin{center}
\subfigure[XT5. The number of cores is fixed at 2,048.]{\label{fig:linear-xt5}
\includegraphics[scale=1,trim=0cm 0cm 0cm 0cm]{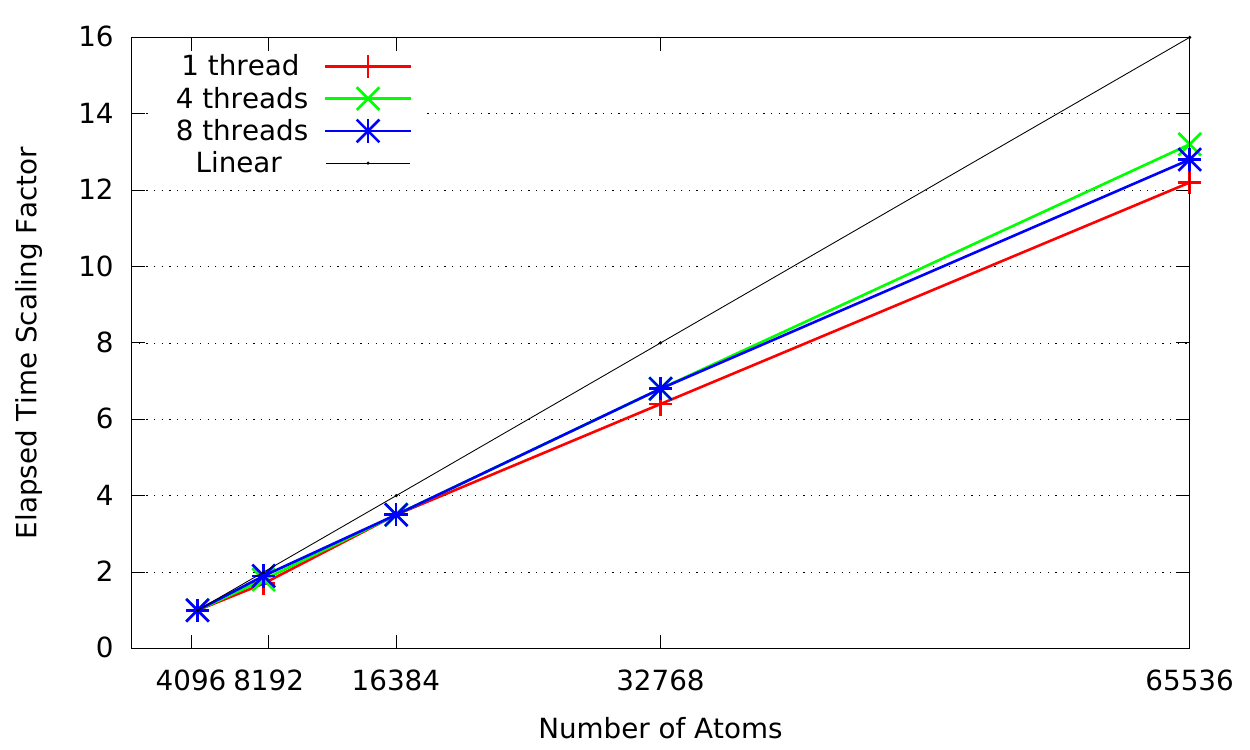}}
\subfigure[K computer. The number of cores is fixed at 16,384.]{\label{fig:linear-k}
\includegraphics[scale=1,trim=0cm 0cm 0cm 0cm]{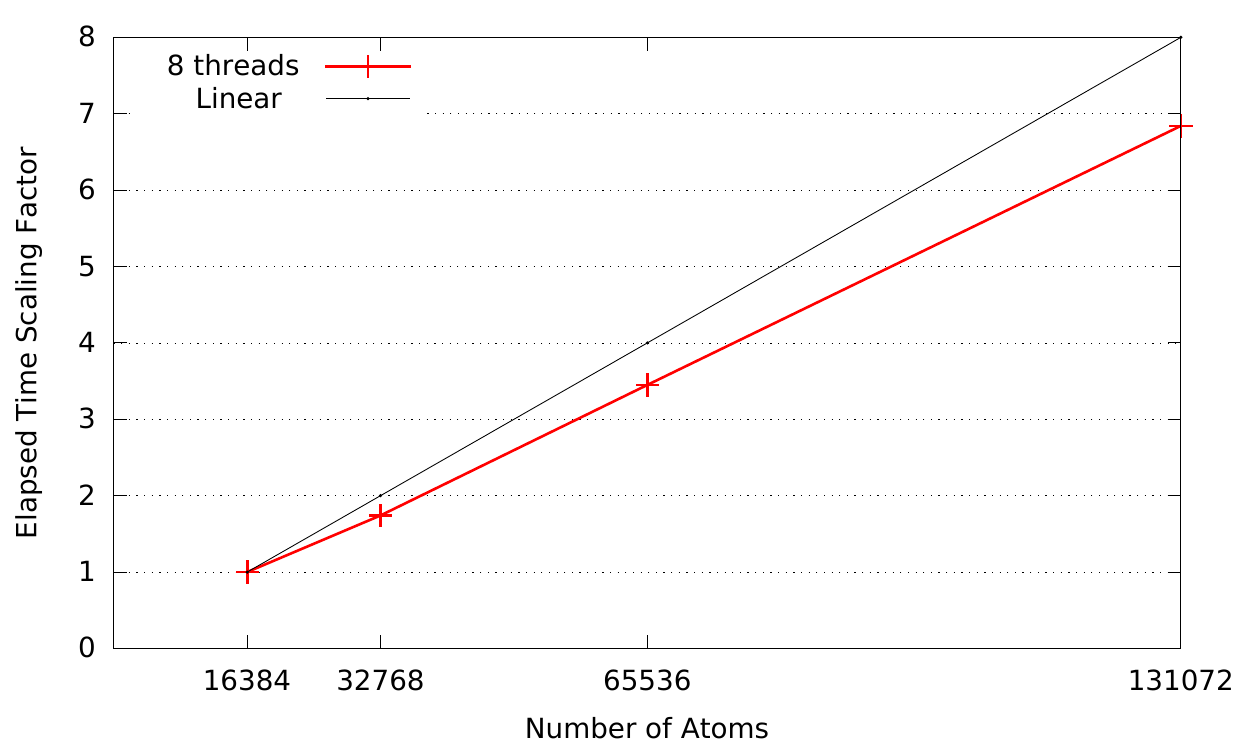}}
\end{center}
\caption{Linear scaling property with the diamond structure.}
\label{linear-xt5-k}
\end{figure}

With the $O(N)$ Krylov subspace method, the elapsed time should be approximately linear to the number of atoms when the number of cores is fixed. Figure \ref{linear-xt5-k} presents the relationship between the elapsed time scaling factor and the number of atoms with a fixed number of cores of 2,048 on XT5 (a), and 16,384 cores on the K computer (b) with the diamond structure. The relationship appears to be nearly linear: an increase in the number of atoms by a factor of 2 leads to an increase in the elapsed time by a factor of a little bit less than 2 in all three cases of 1, 4, and 8 threads. In case of 65,536 atoms in Fig. \ref{linear-xt5-k}(a), the largest and smallest elapsed time scaling factors are 13.2 (4 threads) and 12.1 (1 thread), respectively, as opposed to the linear factor of 16 against the baseline of 4,096 atoms. The same behavior can be observed on the K computer, as shown in Fig. \ref{linear-xt5-k}(b). The reason behind the better-than-linear elapsed time scaling factors is a higher computation to communication ratio with a higher number of atoms per core. When the number of cores is fixed and the number of atoms is increased, the computational amount per core will also be increased, leading to a higher efficiency. For instance, with a fixed number of 16,384 cores on the K computer (Fig. \ref{linear-xt5-k}b), the number of atoms per core in case of 16,384 atoms is only 1, while this number is  8 for 131,072 atoms, resulting in the computational amount per core with 131,072 atoms being 8 times larger than that with 16,384 atoms. Suppose the communication amount stays almost unchanged for a fixed number of cores, the computation to communication ratio with 131,072 atoms would also be 8 times higher than that with 16,384 atoms. Such a linear scaling property is definitely essential to enable extreme-scale applications that require hundreds of thousands to millions of atoms. The property will also be demonstrated with the DNA structure in \ref{strong-DNA}.   

\subsubsection{Strong and weak scaling with the diamond structure}
\begin{figure}[htbp]
\begin{center}
\subfigure[Strong scaling with 16,384 atoms.]{\label{fig:strong-xt5}
\includegraphics[scale=1,trim=0cm 0cm 0cm 0cm]{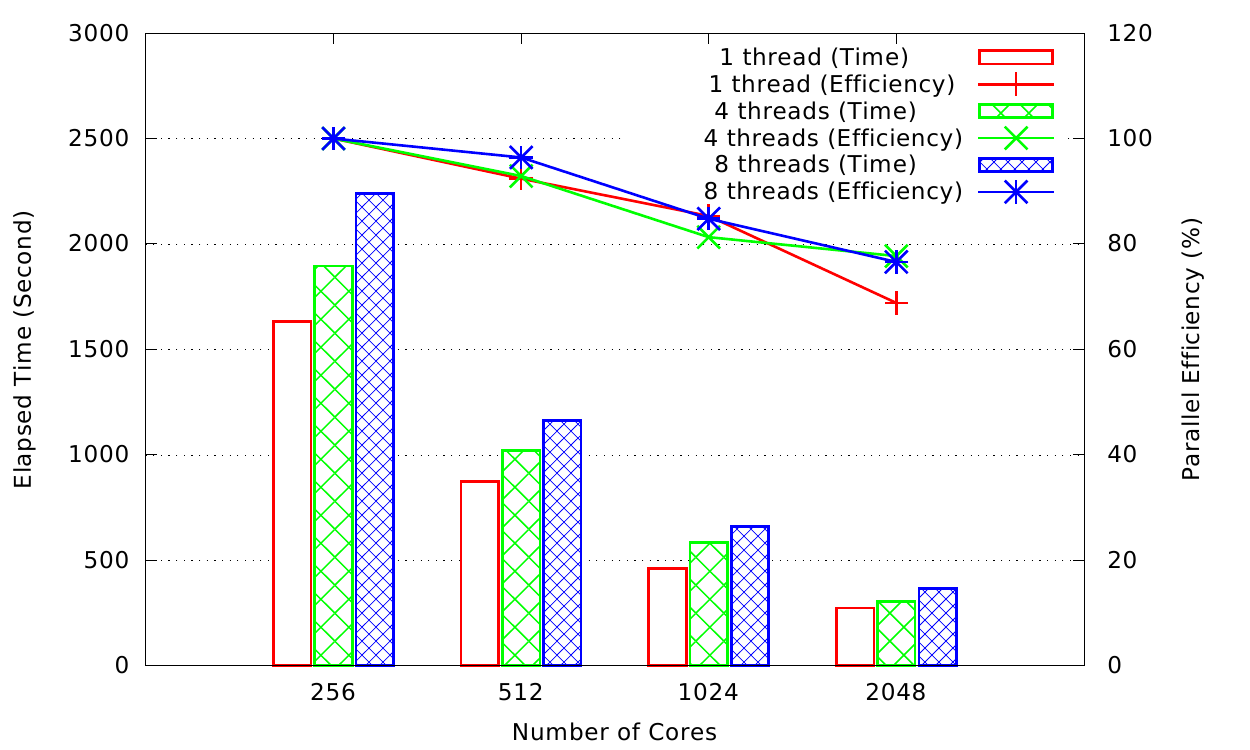}}
\subfigure[Weak scaling with 16 and 32 atoms per core.]{\label{fig:weak-xt5}
\includegraphics[scale=1,trim=0cm 0cm 0cm 0cm]{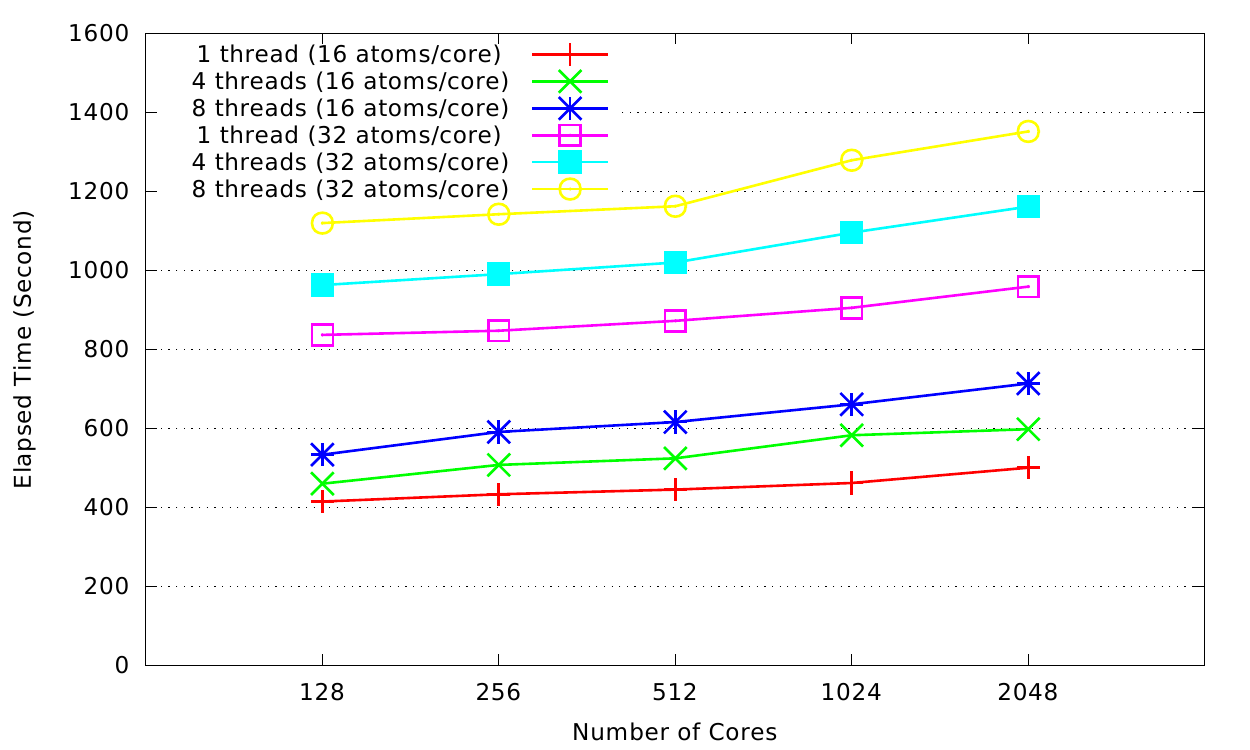}} \\
\end{center}
\caption{(a) Strong and (b) weak scaling on XT5 with the diamond structure.}
\label{strong-weak-xt5}
\end{figure}

\begin{figure}[htbp]
\begin{center}
\subfigure[Strong scaling with 16,384 atoms.]{\label{fig:strong-fx10}
\includegraphics[scale=1,trim=0cm 0cm 0cm 0cm]{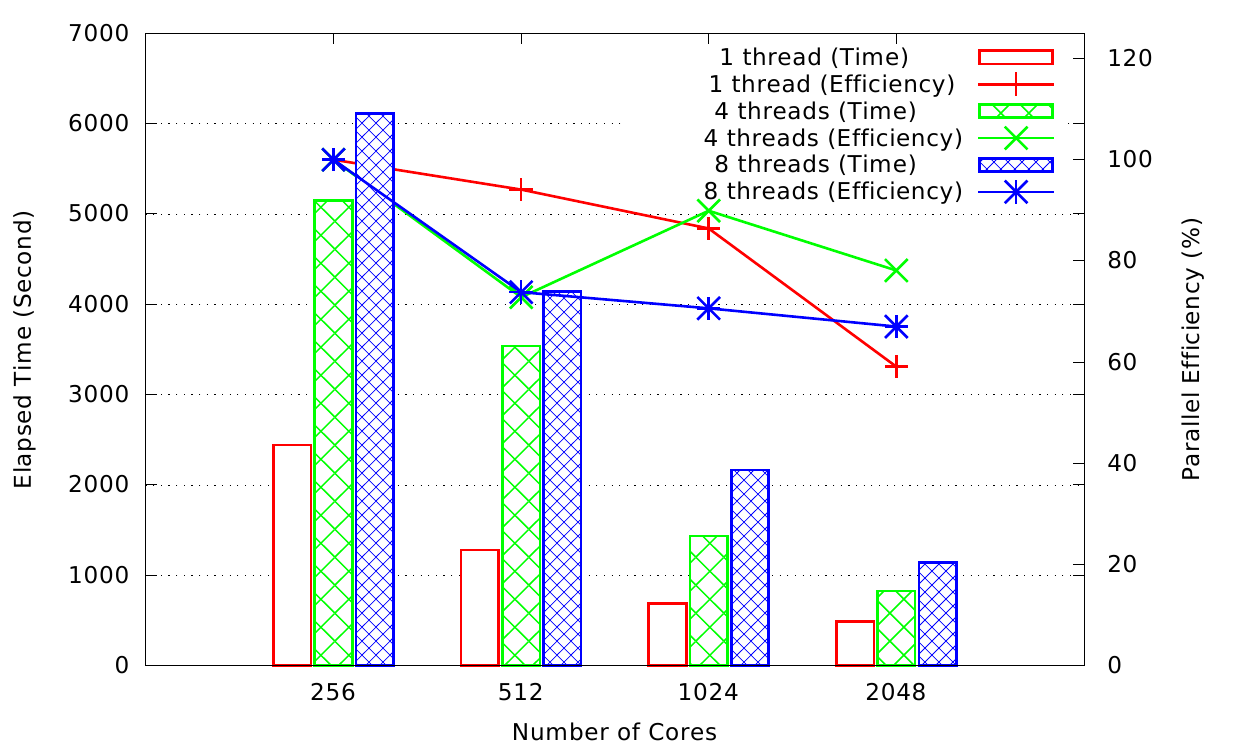}}
\subfigure[Weak scaling with 16 and 32 atoms per core.]{\label{fig:weak-fx10}
\includegraphics[scale=1,trim=0cm 0cm 0cm 0cm]{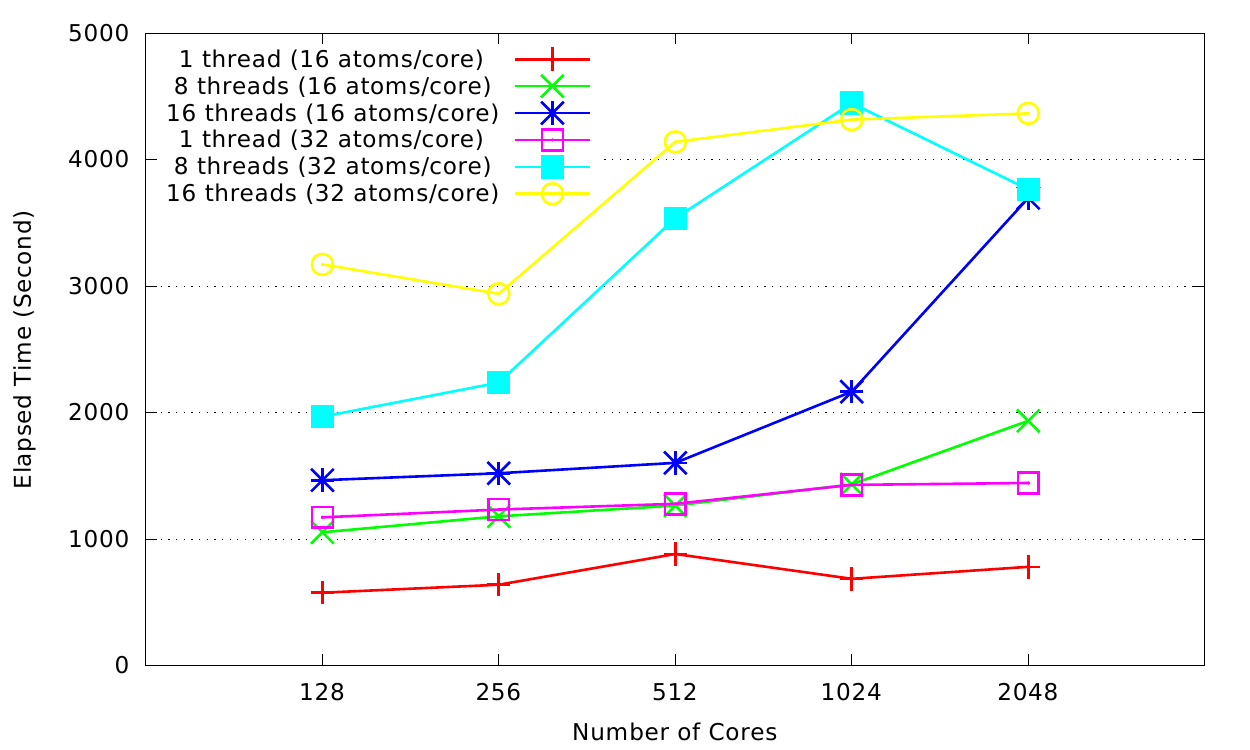}} \\
\end{center}
\caption{(a) Strong and (b) weak scaling on FX10 with the diamond structure.}
\label{strong-weak-fx10}
\end{figure}

\begin{figure}[htbp]
\begin{center}
\subfigure[Strong scaling with 131,072 atoms.]{\label{fig:strong-k}
\includegraphics[scale=1,trim=0cm 0cm 0cm 0cm]{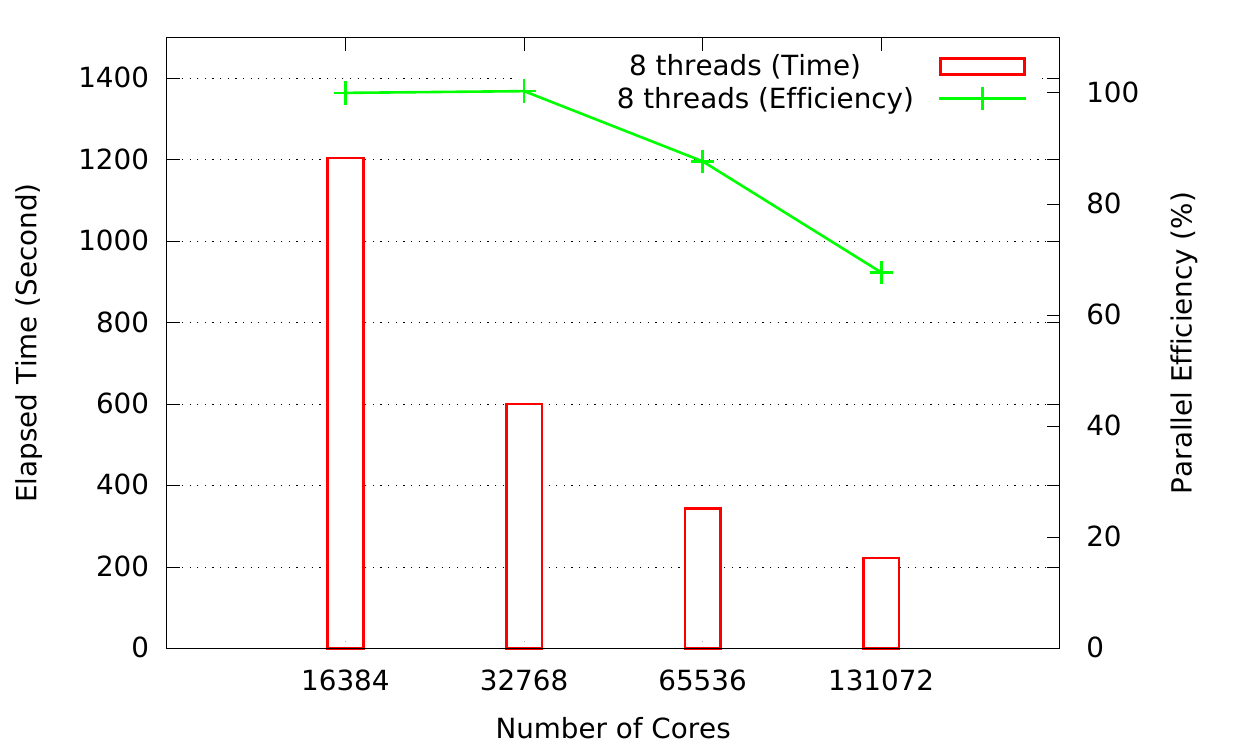}}
\subfigure[Weak scaling with 2 and 4 atoms per core.]{\label{fig:weak-k}
\includegraphics[scale=1,trim=0cm 0cm 0cm 0cm]{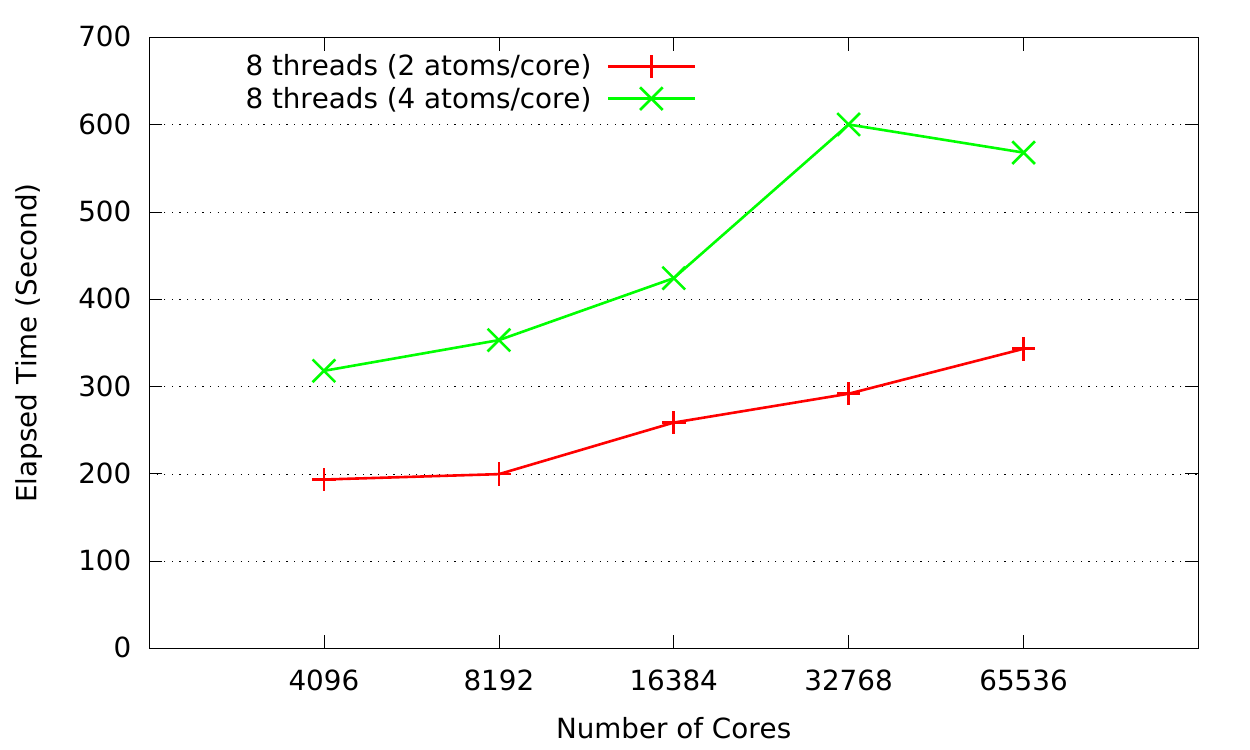}} \\
\end{center}
\caption{(a) Strong and (b) weak scaling on the K computer with the diamond structure.}
\label{strong-weak-k}
\end{figure}

%\begin{figure}[htbp]
%\centering
%\includegraphics[scale=1,trim=0cm 0cm 0cm 0cm]{strong-k-131K.pdf}
%\caption{Strong scaling on the K computer with 131,072 atoms.}
%\label{fig-strong-k}
%\end{figure}

%\begin{figure}[htbp]
%\begin{center}
%\subfigure[Strong Scaling with 131,072 Atoms]{\label{fig:strong-k}
%\includegraphics[scale=1,trim=0cm 0cm 0cm 0cm]{strong-k-131K.pdf}}
%\subfigure[Weak Scaling with 2 and 4 Atoms per Core]{\label{fig:weak-k}
%\includegraphics[scale=1,trim=0cm 0cm 0cm 0cm]{weak-k.pdf}} \\
%\end{center}
%\caption{Strong and weak scaling on the K computer}
%\label{strong-weak-k}
%\end{figure}

Figures \ref{strong-weak-xt5}, \ref{strong-weak-fx10} and \ref{strong-weak-k}, respectively, show the strong scaling property in terms of parallel efficiency and elapsed time, and weak scaling property in terms of elapsed time on XT5, FX10, and the K computer with the diamond structure. In strong scaling calculations, the number of atoms allocated to one processing core is decreased with the increasing number of cores, resulting in a reduction in elapsed time. The strong calculations are performed with 16,384 atoms on both XT5 and FX10 from 256 to 2,048 cores, each with 1, 4, and 8 threads on XT5, and with 1, 8, and 16 threads on FX10. On the K computer, the strong scaling results are taken with 131,072 atoms and up to 131,072 cores in the hybrid mode (8 threads). The parallel efficiency of strong scaling is given by the following equation:
\begin{equation}
\label{eq-parallel-efficiency}
P_\textrm{C}=\frac{\frac{T_\textrm{B}}{T_\textrm{C}}}{\frac{N_\textrm{C}}{N_\textrm{B}}}\times 100\%,
\end{equation}
where $P_\textrm{C}$ is the parallel efficiency of the target calculation, $T_\textrm{B}$ and $N_\textrm{B}$ are the elapsed time and number of cores of the base calculation, respectively, and $T_\textrm{C}$ and $N_\textrm{C}$ are the elapsed time and number of cores of the target calculation, respectively. The base calculation used the equation is the case of 256 cores with XT5 and FX10, and the case of 16,384 cores with the K computer, the lowest numbers of cores. 

On the other hand, weak scaling calculations demand the number of atoms allocated to one processing core be fixed when changing the numbers of cores and atoms to maintain a constant elapsed time in the ideal case, due to a constant computational amount per core. The number of atoms per core in the weak calculations is set at 16 and 32 from 128 cores to 2,048 cores on XT5 and FX10, and 2 and 4 from 4,096 cores to 65,536 cores on the K computer. For example, on XT5 and FX10, in case of 16 atoms per core and 128 cores, there are 2,048 atoms to be processed, and in another case of 2,048 cores and 32 atoms per core, we have to calculate 65,536 atoms. 

We note several observations. First of all, both flat MPI and hybrid MPI/OpenMP modes exhibit good strong scaling property, with the parallel efficiency ranging from 68.8\% to 96.4\% on XT5 (Fig. \ref{strong-weak-xt5}(a)), and from 59.1\% to 94.1\% on FX10 (Fig. \ref{strong-weak-fx10}(a)). On the K computer, the parallel efficiency at 131,072 cores is 67.7\% compared to the baseline of 16,384 cores (Fig. \ref{strong-weak-k}(a)). Also, the parallel efficiency of the hybrid mode tends to be higher than that of the flat MPI mode, especially with the largest number of cores, probably due to a smaller number of MPI processes, accompanied by a smaller amount of communication, for the same number of cores. Using 2,048 cores in the flat MPI mode, the calculation of 16,384 atoms could be finished in less than 250 seconds on XT5.   

For weak scaling on XT5 (Fig. \ref{strong-weak-xt5}(b)), the elapsed time increases gradually with the number of cores due to a growing communication amount caused by more MPI processes. Nevertheless, the increase rate is quite small, ranging from 14.6\% (1 thread with 32 atoms/core) to 33.8\% (8 threads with 16 atoms/core), when the number of cores is raised by a factor of 16 from 128 to 2,048. More importantly, the efficiency maintains its property when the calculation scale is enlarged gradually from 2,048 to 32,768 atoms (16 atoms/core), and from 4,096 to 65,536 atoms (32 atoms/core). This observation suggests that we can expect similar performance for even larger scales, as shown on the K computer (Fig. \ref{strong-weak-k}(b)). On FX10 (Fig. \ref{strong-weak-fx10}(b)), unfortunately we experience some sudden jumps in elapsed time with the hybrid mode, while the flat MPI mode shows a very similar scaling property to that on XT5 with a solid performance (in the same Fig. \ref{strong-weak-fx10}(b)). 

In comparison between the flat MPI and hybrid modes, the elapsed time of the flat MPI mode is always lower than that of the hybrid mode with 4 threads, which in turn is always lower than that of the hybrid mode with 8 threads. The outcome is the same for both strong and weak scaling, asserting that flat MPI is the best mode in both machines at this calculation scale. 

Finally, in weak scaling on XT5 with both hybrid and flat MPI modes and on FX10 with the flat MPI mode, when the number of atoms per core is increased by a factor of 2 from 16 to 32, the elapsed time also increases almost exactly by a factor of 2 as expected. The results again confirm the linear scaling property of OpenMX, and that it can maintain the same level of efficiency for larger calculation scales, as demonstrated on the K computer. 

\subsubsection{Strong scaling property with DNA}
\label{strong-DNA}
\begin{figure}[htbp]
\begin{center}
\subfigure[DNA with 2,600 atoms on the K computer and FX10.]{\label{strong-DNA-4}
\includegraphics[scale=1,trim=0cm 0cm 0cm 0cm]{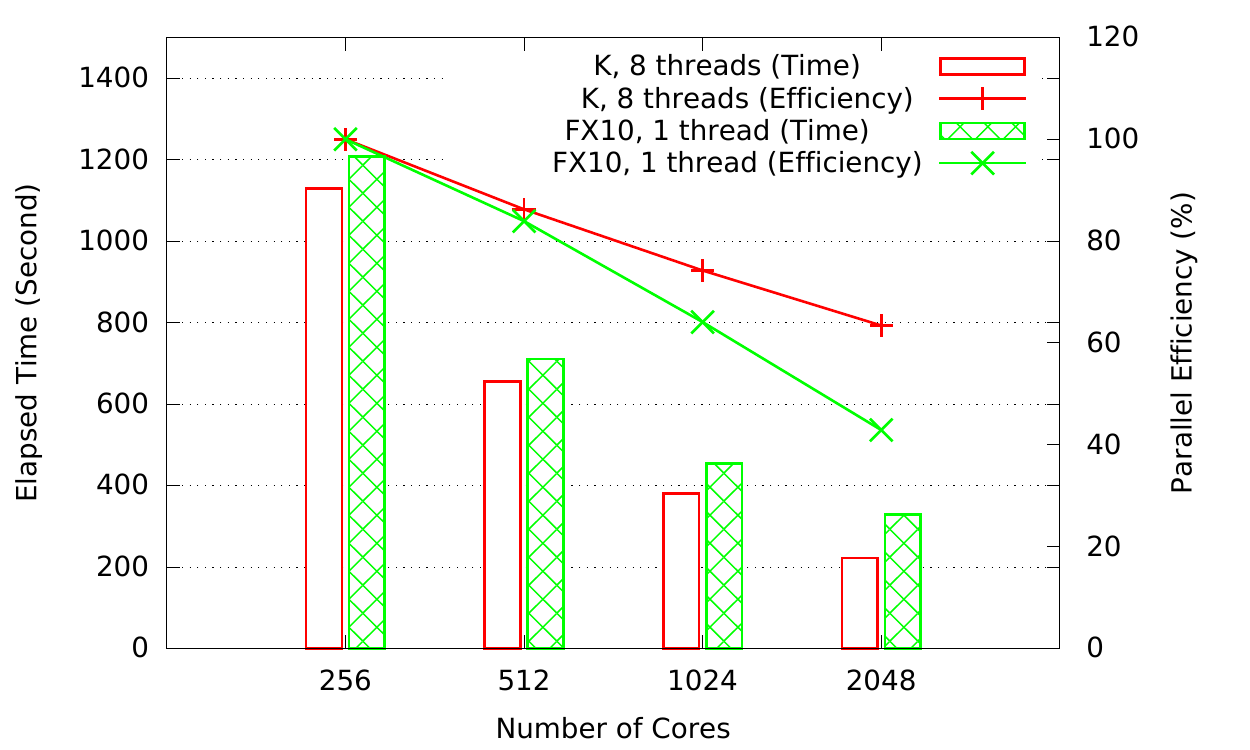}}
\subfigure[DNA with 13,000 and 26,000 atoms on the K computer. ]{\label{strong-DNA-20}
\includegraphics[scale=1,trim=0cm 0cm 0cm 0cm]{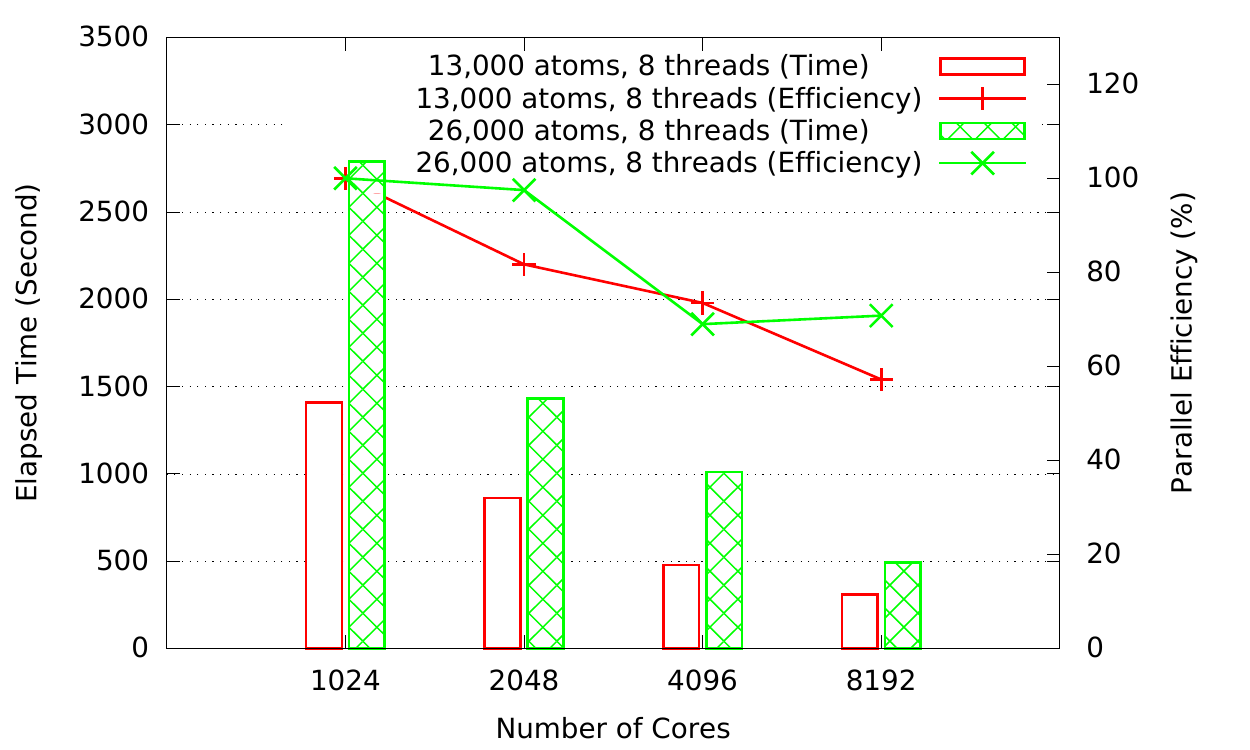}}
\end{center}
\caption{Strong scaling property with the DNA structure.}
\label{strong-scaling-DNA}
\end{figure}

In the diamond case, the structure is uniform, and the numbers of atoms and cores are selected so that each core is assigned the same computational amount. In the case of DNA, by contrast, the structure is long, and the number of atoms is not divisible by the number of cores. Even so, our decomposition scheme is still able to work effectively for the DNA structure, as shown in Fig. \ref{strong-scaling-DNA}, where the parallel efficiency is calculated by Eq. (\ref{eq-parallel-efficiency}) with the baseline being 256 cores in Fig. \ref{strong-scaling-DNA}(a) and 1,024 cores in Fig. \ref{strong-scaling-DNA}(b). With 2,600 atoms and up to 2,048 cores on the K computer (hybrid with 8 threads) and FX10 (flat MPI) (Fig. \ref{strong-scaling-DNA}(a)), the parallel efficiency at 2,048 cores is 63.4\% in the hybrid mode, and drops to only 42.9\% in the flat MPI mode. The efficiency gap is due to a load imbalance caused by the difference in the number of processes in these modes. In the hybrid mode, the number of processes at 2,048 cores is only 256, as against 2,048 in the flat MPI mode. Assume that the atoms have the same weight of 1. Because there are 2,600 atoms, each process is allocated 10 or 11 atoms in the hybrid mode, and 1 or 2 atoms in the flat MPI mode. Obviously, the load imbalance of the flat MPI mode is 100\%, much higher than 10\% of the hybrid mode, resulting in the low efficiency. 

Figure \ref{strong-scaling-DNA}(b) compares the elapsed time and parallel efficiency of 13,000 and 26,000 atoms in the hybrid mode on the K computer with up to 8,192 cores. By observing the elapsed time at each number of cores, we can easily recognize the better-than-linear behavior of our Krylov subspace method, where the elapsed time of 26,000 atoms is nearly twice as long as that of 13,000 atoms. In addition, the parallel efficiency of 26,000 atoms is much higher than that of 13,000 atoms, except for 4,096 cores, for example, 70.8\% compared to 57.2\% at 8,192 cores. The result is attributed to the computational amount per core that is doubled in case of 26,000 atoms, leading to a higher computation to communication ratio and eventually a higher efficiency.  

\subsubsection{Strong scaling property of subroutines}
\begin{figure}[htb]
\centering
\includegraphics[scale=1,trim=0cm 0cm 0cm 0cm]{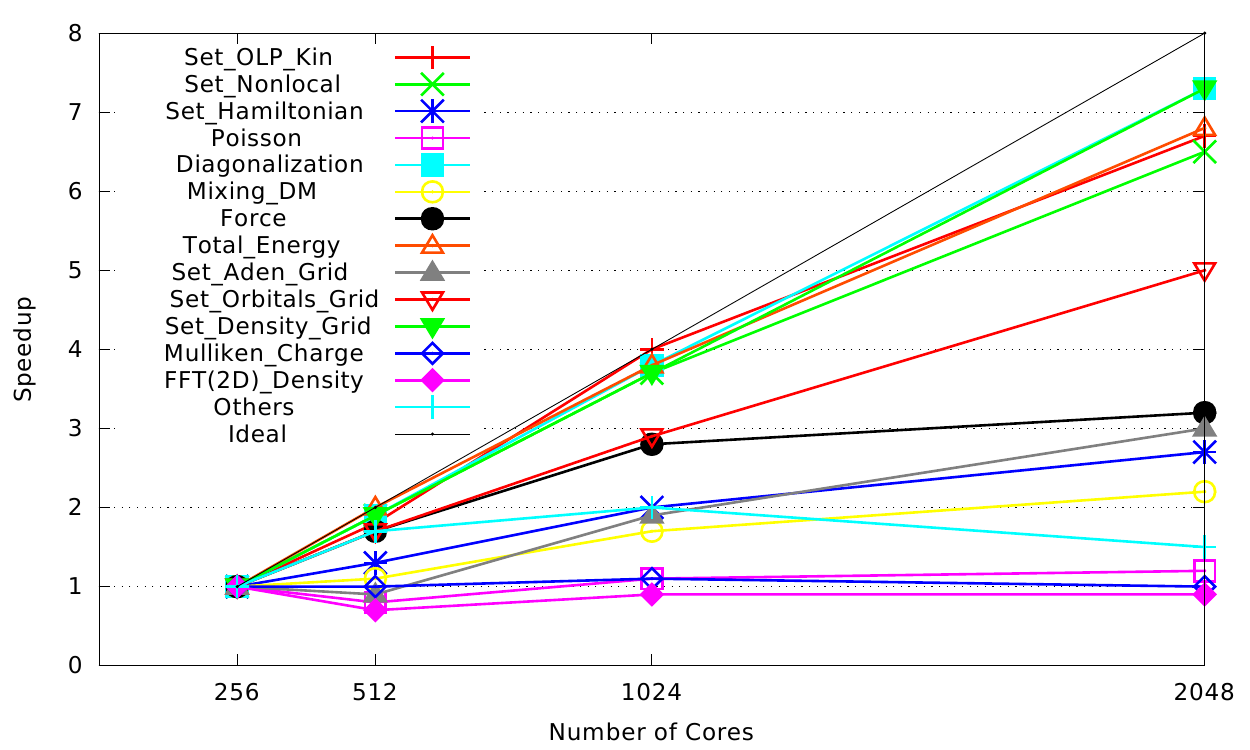}
\caption{Strong scaling property of subroutines. The results are taken with 16,384 diamond atoms, flat MPI on XT5.}
\label{fig-sub-xt5}
\end{figure}

So far we have discussed only the total elapsed time, which should be broken down further to understand the scaling behavior of the subroutines in DFT. Figure \ref{fig-sub-xt5} plots the strong scaling property of the subroutines, taken with 16,384 diamond atoms in the flat MPI mode on XT5. Some important and time-consuming subroutines are noted below in order of descending elapsed time. 

\begin{itemize}
\item
Diagonalization(): is a subroutine for performing the diagonalization, which is actually the O(N) Krylov subspace method in the benchmark series. It accounts for 48.7\% of the total time, and scales well, reaching a speedup of 7.3 at 2,048 cores compared to the ideal speedup of 8. 
\item
Set\_Nonlocal(): is a subroutine for calculating the matrix elements and its derivatives for nonlocal potentials in the momentum space, accounting for 19.6\% of the total time. The speedup at 2,048 cores is 6.5, also good considering that fact that MPI communication is incurred in this subroutine. 
\item
Force(): As the name reveals, this subroutine calculates the force on atoms and contributes 9.4\% to the total time. This is the worst scaling subroutine among the time-consuming ones, as the speedup is only 3.2, less than half of the ideal speedup of 8, caused by a large amount of MPI communication among the processes. This amount can be dramatically reduced if buffer regions are added to store more data, with the tradeoff of incurring high memory usage. Nevertheless, the subroutine is called only once in each MD step, and hence, becoming smaller in calculations that require a large number of SCF iterations. We are now in the process of tuning this subroutine to make it scale better. 
\item
Total\_Energy(): is a subroutine for calculating the total system energy, accounting for 8.3\% of the total time. It exhibits a good scaling property with the speedup of 6.8 at 2,048 cores.   
\item
Set\_OLP\_Kin(): is a subroutine for calculating the overlap matrix and the matrix for the kinetic operator in the momentum space. It accounts for 6.5\% of the total time and also scales well with a speedup of 6.7. 
\item
Poisson(): is a subroutine for solving the Poisson's equation using FFT. Although its contribution to the total time is trivial, approximately 0.02\%, we note it here as it is directly related to our method for grid decomposition.  
\end{itemize}

In summary, all the time-consuming subroutines, except for Force(), which contribute more than 5\% to the total time, scale well with the speedup of 6.5 to 7.3, as against the ideal speedup of 8. As a result, the total time shows a good scaling property as can be seen in previous subsections. 

\subsubsection{Memory usage}
\begin{figure}[htbp]
\centering
\includegraphics[scale=1,trim=0cm 0cm 0cm 0cm]{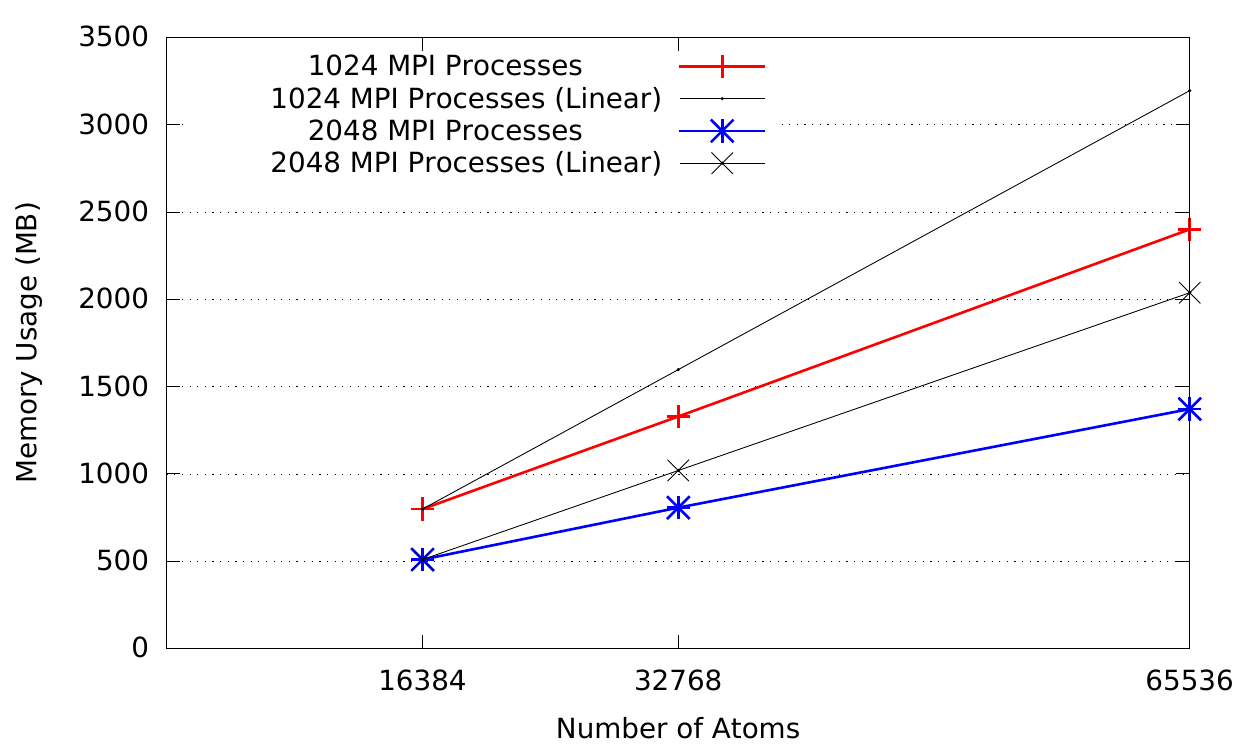}
\caption{Memory usage per process with the diamond structure, measured with 1,024 and 2,048 MPI processes in flat MPI mode on XT5.}
\label{fig-linear-memory}
\end{figure}

Memory management is another crucial aspect that must be handled properly in large-scale calculations. As touched on above, we give priority to efficient use of memory and therefore let the processes carry out on-the-fly communications instead of extending the buffer to store more data in our implementation. Figure \ref{fig-linear-memory} displays the amount of memory per process used in the calculations for 16,384, 32,768, and 65,536 diamond atoms, measured with 1,024 and 2,048 MPI processes in the flat MPI mode on XT5. The numbers of grids of \textit{\textbf{a}}-, \textit{\textbf{b}}-, and \textit{\textbf{c}}-axes are 420, 420, and 210 for 16,384 atoms, 420, 420, and 420 for 32,768 atoms, and 840, 420, and 420 for 65,536 atoms. With the number of processes fixed at 1,024 or 2,048 and the number of atoms increased linearly, the scaling property of memory usage is well below linear, with the scaling factors at 65,536 atoms are 2.7 (1,024 processes) and 3.0 (2,048 processes), as against the linear factor of 4. On the other hand, when the number of processes is doubled from 1,024 to 2,048, the memory usage efficiency factors are 78.4\%, 82.4\%, and 87.6\% for 16,384, 32,768, and 65,536 atoms, respectively.

\section{Conclusion}
\label{Conclusion}
We have developed a three-dimensional domain decomposition scheme that is applicable to both conventional DFT methods and linear scaling DFT methods, consisting of two methods: one for atom decomposition and the other for grid decomposition. The atom decomposition method reorders the atoms along a principal axis based on a modified recursive bisection method and inertia tensor moment. The atoms that are close in real space are also close on the principal axis to maintain data locality. The atoms are then divided into sub-domains in a balanced way among the processes. On the other hand, the grid decomposition method makes it easier for the calculations of charge density and other potentials by defining different data structures for storing the grid data. 
Also, our proposed decomposition method for solving the Poisson equation using FFT in the calculation of Hartree potential exhibits up to 77.8\% enhancement of communication efficiency in comparison to a previously proposed method with $64 \times 64 \times 64$ grid points. 
%We also analyze and employ the best calculation order pattern with communication minimized in this method for calculating the Hartree potential using FFT, which is shown to be up to 77.8\% better than a previously proposed method in terms of communication efficiency. 
Our scheme has been evaluated with OpenMX and the linear scaling Krylov subspace method, and demonstrates good strong and weak scaling properties. On the K computer, the parallel efficiency at 131,072 cores is 67.7\% compared to the baseline of 16,384 cores with 131,072 diamond atoms. The efficiency is from 68.8\% to 96.4\% on XT5, and from 59.1\% to 94.1\% on FX10 with up to 2,048 cores, depending on the number of cores in use. The results suggest that our scheme is efficient for enabling large-scale electronic calculations based on DFT on massively parallel computers.

We are now in the process of tuning the performance of some particular subroutines, especially Force(), the worst scaling subroutine among the time-consuming ones. Should it be improved, certainly we could further accelerate the calculations and achieve a higher parallel efficiency. 



\section*{Acknowledgements}
This work was supported by the Strategic Programs for Innovative Research (SPIRE), MEXT, and the Computational Materials Science Initiative (CMSI), and Materials Design through Computics: Complex Correlation and Non-Equilibrium Dynamics A Grant in Aid for Scientific Research on Innovative Areas, MEXT, Japan. The benchmark calculations were performed using the K computer at RIKEN, Fujitsu FX10 at The University of Tokyo, and Cray XT5 at Japan Advanced Institute of Science and Technology (JAIST). 

%% References


\bibliographystyle{model1a-num-names}
\bibliography{biblio}

\begin{thebibliography}{54}
\expandafter\ifx\csname natexlab\endcsname\relax\def\natexlab#1{#1}\fi
\providecommand{\bibinfo}[2]{#2}
\ifx\xfnm\relax \def\xfnm[#1]{\unskip,\space#1}\fi
%Type = Article
\bibitem[{Hohenberg and Kohn(1964)}]{Hohenberg:1964zz}
\bibinfo{author}{P.~Hohenberg}, \bibinfo{author}{W.~Kohn},
  \bibinfo{journal}{Phys. Rev.} \bibinfo{volume}{136} (\bibinfo{year}{1964})
  \bibinfo{pages}{B864--B871}.
%Type = Article
\bibitem[{Kohn and Sham(1965)}]{Kohn:1965zzb}
\bibinfo{author}{W.~Kohn}, \bibinfo{author}{L.~J. Sham},
  \bibinfo{journal}{Phys. Rev.} \bibinfo{volume}{140} (\bibinfo{year}{1965})
  \bibinfo{pages}{A1133--A1138}.
%Type = Article
\bibitem[{Hedin(1965)}]{Hedin:1965zza}
\bibinfo{author}{L.~Hedin}, \bibinfo{journal}{Phys. Rev.} \bibinfo{volume}{139}
  (\bibinfo{year}{1965}) \bibinfo{pages}{A796--A823}.
%Type = Inproceedings
\bibitem[{Hasegawa et~al.(2011)Hasegawa, Iwata, Tsuji, Takahashi, Oshiyama,
  Minami, Boku, Shoji, Uno, Kurokawa, Inoue, Miyoshi, and
  Yokokawa}]{hasegawa2011first}
\bibinfo{author}{Y.~Hasegawa}, \bibinfo{author}{J.-I. Iwata},
  \bibinfo{author}{M.~Tsuji}, \bibinfo{author}{D.~Takahashi},
  \bibinfo{author}{A.~Oshiyama}, \bibinfo{author}{K.~Minami},
  \bibinfo{author}{T.~Boku}, \bibinfo{author}{F.~Shoji},
  \bibinfo{author}{A.~Uno}, \bibinfo{author}{M.~Kurokawa},
  \bibinfo{author}{H.~Inoue}, \bibinfo{author}{I.~Miyoshi},
  \bibinfo{author}{M.~Yokokawa}, in: \bibinfo{booktitle}{Proceedings of 2011
  International Conference for High Performance Computing, Networking, Storage
  and Analysis}, SC '11, \bibinfo{publisher}{ACM}, \bibinfo{address}{New York,
  NY, USA}, \bibinfo{year}{2011}, pp. \bibinfo{pages}{1:1--1:11}.
%Type = Article
\bibitem[{Saad et~al.(2010)Saad, Chelikowsky, and Shontz}]{saad2010numerical}
\bibinfo{author}{Y.~Saad}, \bibinfo{author}{J.~Chelikowsky},
  \bibinfo{author}{S.~Shontz}, \bibinfo{journal}{SIAM Rev.}
  \bibinfo{volume}{52} (\bibinfo{year}{2010}) \bibinfo{pages}{1}.
%Type = Article
\bibitem[{Tsuchida and Tsukada(1995)}]{PhysRevB.52.5573}
\bibinfo{author}{E.~Tsuchida}, \bibinfo{author}{M.~Tsukada},
  \bibinfo{journal}{Phys. Rev. B} \bibinfo{volume}{52} (\bibinfo{year}{1995})
  \bibinfo{pages}{5573--5578}.
%Type = Article
\bibitem[{Tsuchida and Tsukada(1996)}]{PhysRevB.54.7602}
\bibinfo{author}{E.~Tsuchida}, \bibinfo{author}{M.~Tsukada},
  \bibinfo{journal}{Phys. Rev. B} \bibinfo{volume}{54} (\bibinfo{year}{1996})
  \bibinfo{pages}{7602--7605}.
%Type = Article
\bibitem[{Tsuchida and Tsukada(1998)}]{JPSJ.67.3844}
\bibinfo{author}{E.~Tsuchida}, \bibinfo{author}{M.~Tsukada},
  \bibinfo{journal}{J. Phys. Soc. Jpn.} \bibinfo{volume}{67}
  (\bibinfo{year}{1998}) \bibinfo{pages}{3844--3858}.
%Type = Article
\bibitem[{Pask and Sterne(2005)}]{pask2005finite}
\bibinfo{author}{J.~Pask}, \bibinfo{author}{P.~Sterne},
  \bibinfo{journal}{Model. Simul. Mater. Sci. Eng.} \bibinfo{volume}{13}
  (\bibinfo{year}{2005}) \bibinfo{pages}{R71}.
%Type = Article
\bibitem[{Goedecker(1999)}]{goedecker1999linear}
\bibinfo{author}{S.~Goedecker}, \bibinfo{journal}{Rev. Mod. Phys.}
  \bibinfo{volume}{71} (\bibinfo{year}{1999}) \bibinfo{pages}{1085}.
%Type = Article
\bibitem[{Bowler et~al.(2002)Bowler, Miyazaki, and Gillan}]{bowler2002recent}
\bibinfo{author}{D.~Bowler}, \bibinfo{author}{T.~Miyazaki},
  \bibinfo{author}{M.~Gillan}, \bibinfo{journal}{J. Phys.: Condens. Matter}
  \bibinfo{volume}{14} (\bibinfo{year}{2002}) \bibinfo{pages}{2781}.
%Type = Article
\bibitem[{Goedecker and Scuserza(2003)}]{goedecker2003linear}
\bibinfo{author}{S.~Goedecker}, \bibinfo{author}{G.~Scuserza},
  \bibinfo{journal}{Comp. Sci. Eng.} \bibinfo{volume}{5} (\bibinfo{year}{2003})
  \bibinfo{pages}{14--21}.
%Type = Article
\bibitem[{Bowler and Miyazaki(2012)}]{bowler2012methods}
\bibinfo{author}{D.~Bowler}, \bibinfo{author}{T.~Miyazaki},
  \bibinfo{journal}{Rep. Prog. Phys.} \bibinfo{volume}{75}
  (\bibinfo{year}{2012}) \bibinfo{pages}{036503}.
%Type = Article
\bibitem[{Skylaris et~al.(2005)Skylaris, Haynes, Mostofi, and
  Payne}]{skylaris2005introducing}
\bibinfo{author}{C.~Skylaris}, \bibinfo{author}{P.~Haynes},
  \bibinfo{author}{A.~Mostofi}, \bibinfo{author}{M.~Payne},
  \bibinfo{journal}{J. Chem. Phys.} \bibinfo{volume}{122}
  (\bibinfo{year}{2005}) \bibinfo{pages}{084119}.
%Type = Article
\bibitem[{Yang(1991)}]{yang1991direct}
\bibinfo{author}{W.~Yang}, \bibinfo{journal}{Phys. Rev. Lett.}
  \bibinfo{volume}{66} (\bibinfo{year}{1991}) \bibinfo{pages}{1438--1441}.
%Type = Article
\bibitem[{Yang and Lee(1995)}]{yang1995density}
\bibinfo{author}{W.~Yang}, \bibinfo{author}{T.~Lee}, \bibinfo{journal}{J. Chem.
  Phys.} \bibinfo{volume}{103} (\bibinfo{year}{1995}) \bibinfo{pages}{5674}.
%Type = Article
\bibitem[{Khandogin et~al.(2003)Khandogin, Musier-Forsyth, and
  York}]{khandogin2003insights}
\bibinfo{author}{J.~Khandogin}, \bibinfo{author}{K.~Musier-Forsyth},
  \bibinfo{author}{D.~York}, \bibinfo{journal}{J. Mol. Biol.}
  \bibinfo{volume}{330} (\bibinfo{year}{2003}) \bibinfo{pages}{993--1004}.
%Type = Article
\bibitem[{Ozaki(1999)}]{ozaki1999bond}
\bibinfo{author}{T.~Ozaki}, \bibinfo{journal}{Phys. Rev. B}
  \bibinfo{volume}{59} (\bibinfo{year}{1999}) \bibinfo{pages}{16061--16064}.
%Type = Article
\bibitem[{Ozaki et~al.(2000)Ozaki, Aoki, and Pettifor}]{ozaki2000block}
\bibinfo{author}{T.~Ozaki}, \bibinfo{author}{M.~Aoki}, \bibinfo{author}{D.~G.
  Pettifor}, \bibinfo{journal}{Phys. Rev. B} \bibinfo{volume}{61}
  (\bibinfo{year}{2000}) \bibinfo{pages}{7972--7988}.
%Type = Article
\bibitem[{Ozaki and Terakura(2001)}]{ozaki2001convergent}
\bibinfo{author}{T.~Ozaki}, \bibinfo{author}{K.~Terakura},
  \bibinfo{journal}{Phys. Rev. B} \bibinfo{volume}{64} (\bibinfo{year}{2001})
  \bibinfo{pages}{195126}.
%Type = Article
\bibitem[{Kohn(1996)}]{kohn1996density}
\bibinfo{author}{W.~Kohn}, \bibinfo{journal}{Phys. Rev. Lett.}
  \bibinfo{volume}{76} (\bibinfo{year}{1996}) \bibinfo{pages}{3168--3171}.
%Type = Article
\bibitem[{Stephan and Drabold(1998)}]{stephan1998order}
\bibinfo{author}{U.~Stephan}, \bibinfo{author}{D.~Drabold},
  \bibinfo{journal}{Phys. Rev. B} \bibinfo{volume}{57} (\bibinfo{year}{1998})
  \bibinfo{pages}{6391}.
%Type = Article
\bibitem[{Ordej{\'o}n et~al.(1995)Ordej{\'o}n, Drabold, Martin, and
  Grumbach}]{ordejon1995linear}
\bibinfo{author}{P.~Ordej{\'o}n}, \bibinfo{author}{D.~Drabold},
  \bibinfo{author}{R.~Martin}, \bibinfo{author}{M.~Grumbach},
  \bibinfo{journal}{Phys. Rev. B} \bibinfo{volume}{51} (\bibinfo{year}{1995})
  \bibinfo{pages}{1456}.
%Type = Article
\bibitem[{Mauri et~al.(1993)Mauri, Galli, and Car}]{mauri1993orbital}
\bibinfo{author}{F.~Mauri}, \bibinfo{author}{G.~Galli},
  \bibinfo{author}{R.~Car}, \bibinfo{journal}{Phys. Rev. B}
  \bibinfo{volume}{47} (\bibinfo{year}{1993}) \bibinfo{pages}{9973}.
%Type = Article
\bibitem[{Xiang et~al.(2006)Xiang, Li, Liang, Yang, Hou, and
  Zhu}]{xiang2006linear}
\bibinfo{author}{H.~Xiang}, \bibinfo{author}{Z.~Li},
  \bibinfo{author}{W.~Liang}, \bibinfo{author}{J.~Yang},
  \bibinfo{author}{J.~Hou}, \bibinfo{author}{Q.~Zhu}, \bibinfo{journal}{J.
  Chem. Phys.} \bibinfo{volume}{124} (\bibinfo{year}{2006})
  \bibinfo{pages}{234108}.
%Type = Article
\bibitem[{Rudberg and Rubensson(2011)}]{rudberg2011assessment}
\bibinfo{author}{E.~Rudberg}, \bibinfo{author}{E.~Rubensson},
  \bibinfo{journal}{J. Phys.: Condens. Matter} \bibinfo{volume}{23}
  (\bibinfo{year}{2011}) \bibinfo{pages}{075502}.
%Type = Article
\bibitem[{Jordan and Mazziotti(2005)}]{jordan2005comparison}
\bibinfo{author}{D.~Jordan}, \bibinfo{author}{D.~Mazziotti},
  \bibinfo{journal}{J. Chem. Phys.} \bibinfo{volume}{122}
  (\bibinfo{year}{2005}) \bibinfo{pages}{084114}.
%Type = Article
\bibitem[{Daw(1993)}]{PhysRevB.47.10895}
\bibinfo{author}{M.~S. Daw}, \bibinfo{journal}{Phys. Rev. B}
  \bibinfo{volume}{47} (\bibinfo{year}{1993}) \bibinfo{pages}{10895--10898}.
%Type = Article
\bibitem[{Li et~al.(1993)Li, Nunes, and Vanderbilt}]{PhysRevB.47.10891}
\bibinfo{author}{X.-P. Li}, \bibinfo{author}{R.~W. Nunes},
  \bibinfo{author}{D.~Vanderbilt}, \bibinfo{journal}{Phys. Rev. B}
  \bibinfo{volume}{47} (\bibinfo{year}{1993}) \bibinfo{pages}{10891--10894}.
%Type = Article
\bibitem[{Goedecker and Colombo(1994)}]{goedecker1994efficient}
\bibinfo{author}{S.~Goedecker}, \bibinfo{author}{L.~Colombo},
  \bibinfo{journal}{Phys. Rev. Lett.} \bibinfo{volume}{73}
  (\bibinfo{year}{1994}) \bibinfo{pages}{122--125}.
%Type = Article
\bibitem[{Goedecker and Teter(1995)}]{goedecker1995tight}
\bibinfo{author}{S.~Goedecker}, \bibinfo{author}{M.~Teter},
  \bibinfo{journal}{Phys. Rev. B} \bibinfo{volume}{51} (\bibinfo{year}{1995})
  \bibinfo{pages}{9455--9464}.
%Type = Article
\bibitem[{Krajewski and Parrinello(2005)}]{krajewski2005stochastic}
\bibinfo{author}{F.~R. Krajewski}, \bibinfo{author}{M.~Parrinello},
  \bibinfo{journal}{Phys. Rev. B} \bibinfo{volume}{71} (\bibinfo{year}{2005})
  \bibinfo{pages}{233105}.
%Type = Article
\bibitem[{Krajewski and Parrinello(2006)}]{krajewski2006linear}
\bibinfo{author}{F.~R. Krajewski}, \bibinfo{author}{M.~Parrinello},
  \bibinfo{journal}{Phys. Rev. B} \bibinfo{volume}{73} (\bibinfo{year}{2006})
  \bibinfo{pages}{041105}.
%Type = Article
\bibitem[{Ozaki(2006)}]{ozaki2006n}
\bibinfo{author}{T.~Ozaki}, \bibinfo{journal}{Phys. Rev. B}
  \bibinfo{volume}{74} (\bibinfo{year}{2006}) \bibinfo{pages}{245101}.
%Type = Article
\bibitem[{Bowler and Miyazaki(2010)}]{bowler2010calculations}
\bibinfo{author}{D.~Bowler}, \bibinfo{author}{T.~Miyazaki},
  \bibinfo{journal}{J. Phys.: Condens. Matter} \bibinfo{volume}{22}
  (\bibinfo{year}{2010}) \bibinfo{pages}{074207}.
%Type = Article
\bibitem[{Artacho et~al.(2008)Artacho, Anglada, Di{\'e}guez, Gale, Garc{\'\i}a,
  Junquera, Martin, Ordej{\'o}n, Pruneda, S{\'a}nchez-Portal
  et~al.}]{artacho2008siesta}
\bibinfo{author}{E.~Artacho}, \bibinfo{author}{E.~Anglada},
  \bibinfo{author}{O.~Di{\'e}guez}, \bibinfo{author}{J.~Gale},
  \bibinfo{author}{A.~Garc{\'\i}a}, \bibinfo{author}{J.~Junquera},
  \bibinfo{author}{R.~Martin}, \bibinfo{author}{P.~Ordej{\'o}n},
  \bibinfo{author}{J.~Pruneda}, \bibinfo{author}{D.~S{\'a}nchez-Portal},
  et~al., \bibinfo{journal}{J. Phys.: Condens. Matter} \bibinfo{volume}{20}
  (\bibinfo{year}{2008}) \bibinfo{pages}{064208}.
%Type = Article
\bibitem[{Haynes et~al.(2006)Haynes, Skylaris, Mostofi, and
  Payne}]{haynes2006onetep}
\bibinfo{author}{P.~Haynes}, \bibinfo{author}{C.~Skylaris},
  \bibinfo{author}{A.~Mostofi}, \bibinfo{author}{M.~Payne},
  \bibinfo{journal}{Phys. Stat. Sol. B} \bibinfo{volume}{243}
  (\bibinfo{year}{2006}) \bibinfo{pages}{2489--2499}.
%Type = Article
\bibitem[{Br{\'a}zdov{\'a} and Bowler(2008)}]{brazdova2008automatic}
\bibinfo{author}{V.~Br{\'a}zdov{\'a}}, \bibinfo{author}{D.~Bowler},
  \bibinfo{journal}{J. Phys.: Condens. Matter} \bibinfo{volume}{20}
  (\bibinfo{year}{2008}) \bibinfo{pages}{275223}.
%Type = Article
\bibitem[{Challacombe(2000)}]{challacombe2000general}
\bibinfo{author}{M.~Challacombe}, \bibinfo{journal}{Comp. Phys. Comm.}
  \bibinfo{volume}{128} (\bibinfo{year}{2000}) \bibinfo{pages}{93--107}.
%Type = Book
\bibitem[{Hehre et~al.(1986)Hehre, Radom, Schleyer, Pople et~al.}]{hehre1986ab}
\bibinfo{author}{W.~Hehre}, \bibinfo{author}{L.~Radom},
  \bibinfo{author}{P.~Schleyer}, \bibinfo{author}{J.~Pople}, et~al.,
  \bibinfo{title}{Ab initio molecular orbital theory},
  volume~\bibinfo{volume}{9}, \bibinfo{publisher}{Wiley, New York},
  \bibinfo{year}{1986}.
%Type = Article
\bibitem[{S{\'a}nchez-Portal et~al.(1997)S{\'a}nchez-Portal, Ordejon, Artacho,
  and Soler}]{sanchez1997density}
\bibinfo{author}{D.~S{\'a}nchez-Portal}, \bibinfo{author}{P.~Ordejon},
  \bibinfo{author}{E.~Artacho}, \bibinfo{author}{J.~Soler},
  \bibinfo{journal}{Int. J. Quant. Chem.} \bibinfo{volume}{65}
  (\bibinfo{year}{1997}) \bibinfo{pages}{453--461}.
%Type = Article
\bibitem[{Ozaki(2003)}]{ozaki2003variationally}
\bibinfo{author}{T.~Ozaki}, \bibinfo{journal}{Phys. Rev. B}
  \bibinfo{volume}{67} (\bibinfo{year}{2003}) \bibinfo{pages}{155108}.
%Type = Inproceedings
\bibitem[{Thomas(1927)}]{thomas1927calculation}
\bibinfo{author}{L.~Thomas}, in: \bibinfo{booktitle}{Proc. Camb. Phil. Soc.},
  volume~\bibinfo{volume}{23}, \bibinfo{organization}{Cambridge Univ Press},
  pp. \bibinfo{pages}{542--548}.
%Type = Article
\bibitem[{Slater(1951)}]{slater1951simplification}
\bibinfo{author}{J.~Slater}, \bibinfo{journal}{Phys. Rev.} \bibinfo{volume}{81}
  (\bibinfo{year}{1951}) \bibinfo{pages}{385}.
%Type = Book
\bibitem[{Parr and Yang(1994)}]{parr1994density}
\bibinfo{author}{R.~Parr}, \bibinfo{author}{W.~Yang},
  \bibinfo{title}{Density-functional theory of atoms and molecules},
  volume~\bibinfo{volume}{16}, \bibinfo{publisher}{Oxford Univ. Pr.},
  \bibinfo{year}{1994}.
%Type = Book
\bibitem[{Martin(2004)}]{martin2004electronic}
\bibinfo{author}{R.~Martin}, \bibinfo{title}{Electronic structure: basic theory
  and practical methods}, \bibinfo{publisher}{Cambridge Univ. Pr.},
  \bibinfo{year}{2004}.
%Type = Article
\bibitem[{Ohwaki et~al.(2012)Ohwaki, Otani, Ikeshoji, and
  Ozaki}]{ohwaki2012large}
\bibinfo{author}{T.~Ohwaki}, \bibinfo{author}{M.~Otani},
  \bibinfo{author}{T.~Ikeshoji}, \bibinfo{author}{T.~Ozaki},
  \bibinfo{journal}{J. Chem. Phys.} \bibinfo{volume}{136}
  (\bibinfo{year}{2012}) \bibinfo{pages}{134101}.
%Type = Article
\bibitem[{Sawada et~al.(2012)Sawada, Taniguchi, Kawakami, and
  Ozaki}]{Sawada2012}
\bibinfo{author}{H.~Sawada}, \bibinfo{author}{S.~Taniguchi},
  \bibinfo{author}{K.~Kawakami}, \bibinfo{author}{T.~Ozaki},
  \bibinfo{journal}{submitted}  (\bibinfo{year}{2012}).
%Type = Book
\bibitem[{Salmon(1991)}]{salmon1991parallel}
\bibinfo{author}{J.~Salmon}, \bibinfo{title}{Parallel hierarchical N-body
  methods}, \bibinfo{publisher}{Physics, Mathematics and Astronomy Pub.,
  California Institute of Technology}, \bibinfo{year}{1991}.
%Type = Article
\bibitem[{Ozaki and Kino(2005)}]{PhysRevB.72.045121}
\bibinfo{author}{T.~Ozaki}, \bibinfo{author}{H.~Kino}, \bibinfo{journal}{Phys.
  Rev. B} \bibinfo{volume}{72} (\bibinfo{year}{2005}) \bibinfo{pages}{045121}.
%Type = Article
\bibitem[{Perdew et~al.(1996)Perdew, Burke, and
  Ernzerhof}]{perdew1996generalized}
\bibinfo{author}{J.~Perdew}, \bibinfo{author}{K.~Burke},
  \bibinfo{author}{M.~Ernzerhof}, \bibinfo{journal}{Phys. Rev. Lett.}
  \bibinfo{volume}{77} (\bibinfo{year}{1996}) \bibinfo{pages}{3865--3868}.
%Type = Article
\bibitem[{Perdew and Zunger(1981)}]{perdew1981self}
\bibinfo{author}{J.~Perdew}, \bibinfo{author}{A.~Zunger},
  \bibinfo{journal}{Phys. Rev. B} \bibinfo{volume}{23} (\bibinfo{year}{1981})
  \bibinfo{pages}{5048}.
%Type = Incollection
\bibitem[{Takahashi(2010)}]{takahashi2010implementation}
\bibinfo{author}{D.~Takahashi}, in: \bibinfo{editor}{R.~Wyrzykowski},
  \bibinfo{editor}{J.~Dongarra}, \bibinfo{editor}{K.~Karczewski},
  \bibinfo{editor}{J.~Wasniewski} (Eds.), \bibinfo{booktitle}{Parallel
  Processing and Applied Mathematics}, volume \bibinfo{volume}{6067} of
  \textit{\bibinfo{series}{Lecture Notes in Computer Science}},
  \bibinfo{publisher}{Springer Berlin / Heidelberg}, \bibinfo{year}{2010}, pp.
  \bibinfo{pages}{606--614}.
%Type = Misc
\bibitem[{OpenMX(8 20)}]{openmx}
\bibinfo{author}{OpenMX}, \bibinfo{title}{Open source package for {M}aterial
  e{X}plorer}, \bibinfo{howpublished}{{http://www.openmx-square.org/}},
  \bibinfo{year}{retrieved 2012-08-20}.

\end{thebibliography}






% \end{thebibliography}

\end{document}